\documentclass[12pt,notitlepage]{article}

\usepackage{xcolor}
\usepackage{graphicx}
\usepackage{wrapfig}
\usepackage{indentfirst}
\usepackage{titling}
\usepackage{appendix}
\usepackage[most]{tcolorbox}
\usepackage{dcolumn}
\newcolumntype{.}{D{.}{.}{-1}}
\usepackage[super]{nth}
\usepackage{booktabs,caption}
\usepackage[flushleft]{threeparttable}
\usepackage{subcaption}
\usepackage{hyperref}
\usepackage{makecell}
\usepackage{rotating}
\usepackage{capt-of}
\usepackage{wrapfig} 
\usepackage{amsfonts}
\usepackage{pdflscape}
\usepackage{longtable}
\usepackage{booktabs}
\usepackage{array}
\usepackage[authordate,backend=biber,natbib,noibid]{biblatex-chicago}
\renewenvironment{abstract}
 {\small
  \begin{center}
  \bfseries \abstractname\vspace{-.5em}\vspace{0pt}
  \end{center}
  \list{}{%
    \setlength{\leftmargin}{0mm}
    \setlength{\rightmargin}{\leftmargin}%
  }%
  \item\relax}
 {\endlist}
\usepackage{varioref}
\usepackage{hyperref}
\usepackage{cleveref}
\hypersetup{
    colorlinks=true,
    allcolors = gray
    }
\usepackage{titling}
\title{Generative AI at the Crossroads:\\Light Bulb, Dynamo, or Microscope?\thanks{Authors are listed in alphabetical order, not in order of relative contribution. Baily and Kane are at the Brookings Institution (\mbox{mbaily@brookings.edu} and \mbox{akane@brookings.edu}). Byrne and Soto are at the Federal Reserve Board of Governors (\mbox{david.m.byrne@frb.gov} and \mbox{paul.e.soto@frb.gov}).  The views expressed here are not represented to be the views of the staff or trustees of The Brookings Institution nor of the Federal Reserve.  The authors are grateful to Michael Chui, Leland Crane, Avi Goldfarb, Bob Gordon, Shane Greenstein, Anton Korinek, James Manyika, Sid Srinivasan, Scott Stern, and Bill Whyman for helpful conversations.}}
\date{\today \vspace{-2em} }
\author{\begin{tabular}{cc}
Martin Neil Baily &\hspace{2em} David M. Byrne \\  Aidan T. Kane &\hspace{2em} Paul E. Soto\end{tabular}}
\begin{document}
\maketitle
\begin{abstract}With the advent of generative AI (genAI), the potential scope of artificial intelligence has increased dramatically, but the future effect of genAI on productivity remains uncertain. The effect of the technology on the innovation process is a crucial open question.  Some inventions, such as the light bulb, temporarily raise productivity growth as adoption spreads, but the effect fades when the market is saturated; that is, the level of output per hour is permanently higher but the growth rate is not.  In contrast, two types of technologies stand out as having longer-lived effects on productivity growth.  First, there are technologies known as general-purpose technologies (GPTs).  GPTs (1) are widely adopted, (2) spur abundant knock-on innovations (new goods and services, process efficiencies, and business reorganization), and (3) show continual improvement, refreshing this innovation cycle; the electric dynamo is an example.  Second, there are inventions of methods of invention (IMIs). IMIs increase the efficiency of the research and development process via improvements to observation, analysis, communication, or organization; the compound microscope is an example. We show that GenAI has the characteristics of both a GPT and an IMI---an encouraging sign that genAI will raise the \textit{level} of productivity.  Even so, genAI's contribution to productivity \textit{growth} will depend on the speed with which that level is attained and, historically, integrating revolutionary technologies into the economy is a protracted process.\end{abstract}

\section{Introduction}

\begin{figure}[ht]\caption{\label{figure:ai-hai-human-level-performance}AI Benchmark Performance}
	\centering
	\includegraphics[width=\linewidth]{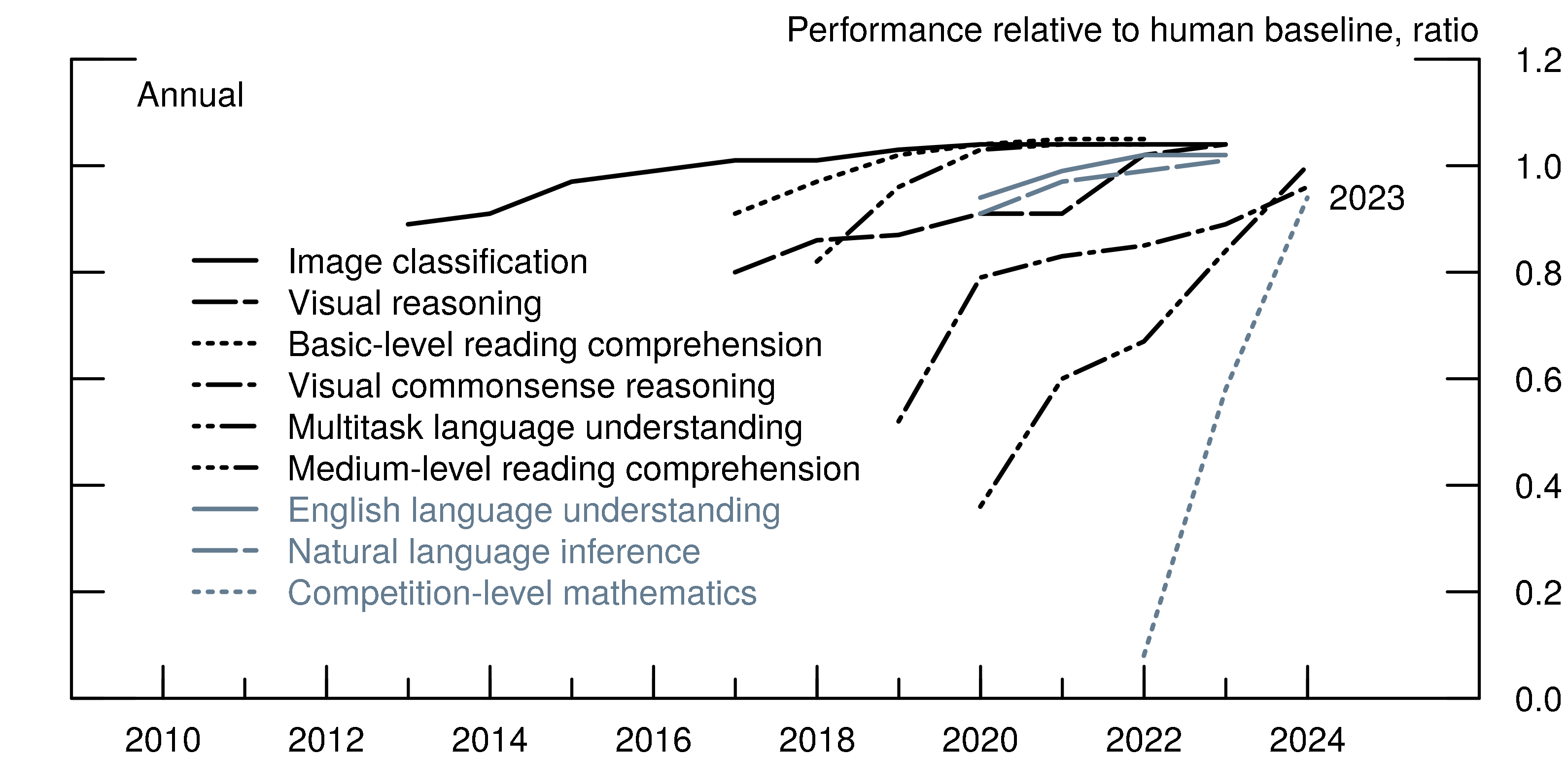}
\begin{minipage}{\linewidth}\setlength{\belowdisplayskip}{0pt} \setlength{\abovedisplayskip}{0pt}
\scriptsize
		{\vspace{2em}Note: The ``human baseline'' concept used varies by task.  For more challenging tasks, the baseline tends to reflect expert-level performance.\\Source: Reproduced with permission from the 2024 AI Index Report, Stanford Institute for Human-centered Artificial Intelligence.}
\end{minipage}
\end{figure}

In late 2022, OpenAI grabbed the world's attention with ChatGPT, the first of several recently released ``generative AI'' (genAI) programs that use a computer model of human discourse to respond to natural-language questions. The scope of AI has expanded dramatically with the advent of genAI, including to tasks previously seen as quintessentially human, such as competition-level mathematics (\vref{figure:ai-hai-human-level-performance}).   Indeed, more and more challenging benchmark tests have been needed to assess progress as genAI has matched human performance on one task after another.\footnote{For more on the record of AI benchmark performance, see \citet{AI-Index-2024}. The external validity of such benchmarks---that is, how much they tell us about performance on practical tasks seemingly related to the tests---is a matter of some debate \citep{liao-taori-inioluwa-schmidt:2021:learning-yet-meta-review-evaluation}.  New benchmarks, such as the ARC-AGI and the Graduate-Level Google-Proof Q\&A (GPQA) benchmarks, have been introduced to better evaluate advanced capabilities. Recent models, especially ``reasoning" models such as o3 from OpenAI and others discussed in \vref{section:model-development}, have performed strongly on these more demanding tasks.  \citet{haupt-brynjolfsson:2025:ai-imitation-game-centaur-evaluations} argue that benchmarks that measure how well AI and humans can jointly perform tasks are needed to shed light on practical AI use.}  In an encouraging sign, field test evidence of productivity improvements from genAI in practical applications has also emerged, including for writing, computer programming, and responding to call center inquiries (\vref{table:genai-field-studies}).\footnote{\citet{brynjolfsson-li-raymond:2025:generative-ai-work} find that call center operators became 14\% more productive when they used the technology; \citet{peng-kalliamvakou-cihon-demirer:2023:impact-ai-developer-productivity-evidence} find that access to GitHub Copilot enabled programmers to complete tasks almost 56\% faster; and \citet{noy-zhang:2023:experimental-evidence-productivity-effects-generative} find that writers who used ChatGPT worked more quickly and produced higher-quality outputs.}  It remains to be seen whether widely-used cost-effective business applications will follow from these successful field tests.  Although some companies do credit genAI with improvement to their bottom line, \citet{mckinsey:2025:state-ai-2025} reports that more than 80\% of genAI-using firms ``aren't seeing a tangible impact on enterprise-level [earnings before interest and taxes] from their use of genAI.''\footnote{\citet{mckinsey:2025:state-ai-2025} also reports that a large share of their survey respondents---primarily large corporations---credit AI with cost reductions in some business functions.} 
%

\begin{table}
    \centering
    \caption{Selected GenAI Productivity Field Studies}
    \label{table:genai-field-studies}
    \begin{tabular}{|p{4.5cm}|p{2cm}|p{6cm}|}
    \hline
        Study & Task & Results \\ \hline \hline
        Noy and Zhang (2023) & Writing & \mbox{ChatGPT} speeds, improves writing. Writers shift from drafting to idea generation and editing.\\ \hline
        \mbox{Brynjolfsson et al.,} \mbox{(2023)} & Customer \mbox{service} & More issues resolved using conversational assistant. \\ \hline
        Dell'Aqua et al. (2023) & Various & GPT-4 increases task completion, speed, and quality.\\ \hline
        Peng et al. (2023) & Coding & Using GitHub Copilot, programmers complete tasks faster \\ \hline        Cui et al. (2024) \nocite{cui-etal:2024:productivity-effects-generative-ai-evidence}& Coding & GitHub Copilot raises task completion.\\ \hline
    \end{tabular}
    \label{table:selected-genai-studies}
\end{table}

\begin{figure}[ht]

\centering
\caption{Indicators of Interest in GenAI}\label{figure:indicators-interest-genai}
\begin{subfigure}[t]{0.48\textwidth}
 \caption{\label{figure:genai-downloads}GenAI Mobile App Downloads}
    \includegraphics[width=\textwidth]{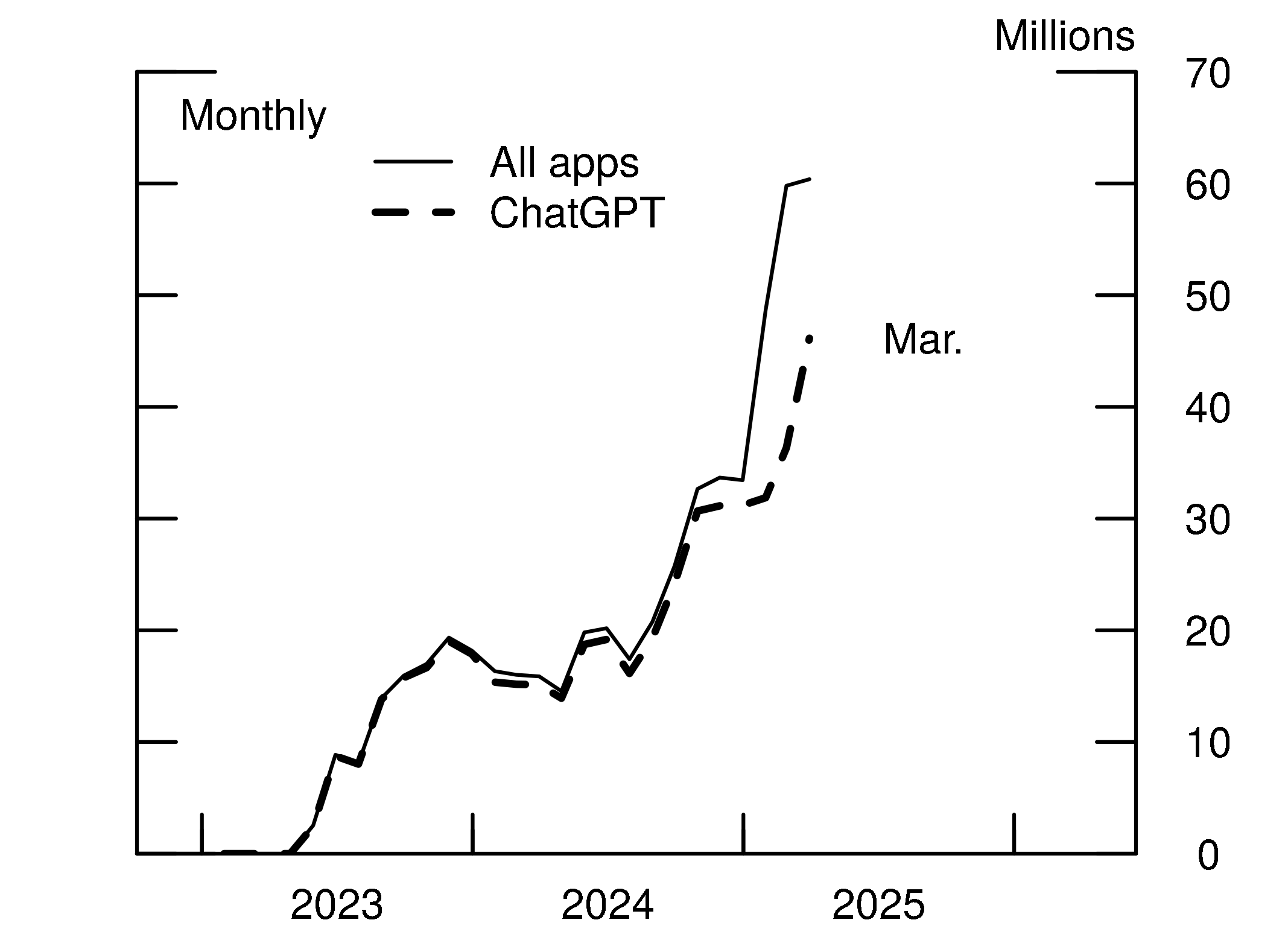}
\end{subfigure}
\begin{subfigure}[t]{0.48\textwidth}
\caption{\label{figure:google-trends_ai}Web Searches for AI}   
\includegraphics[width=\textwidth]{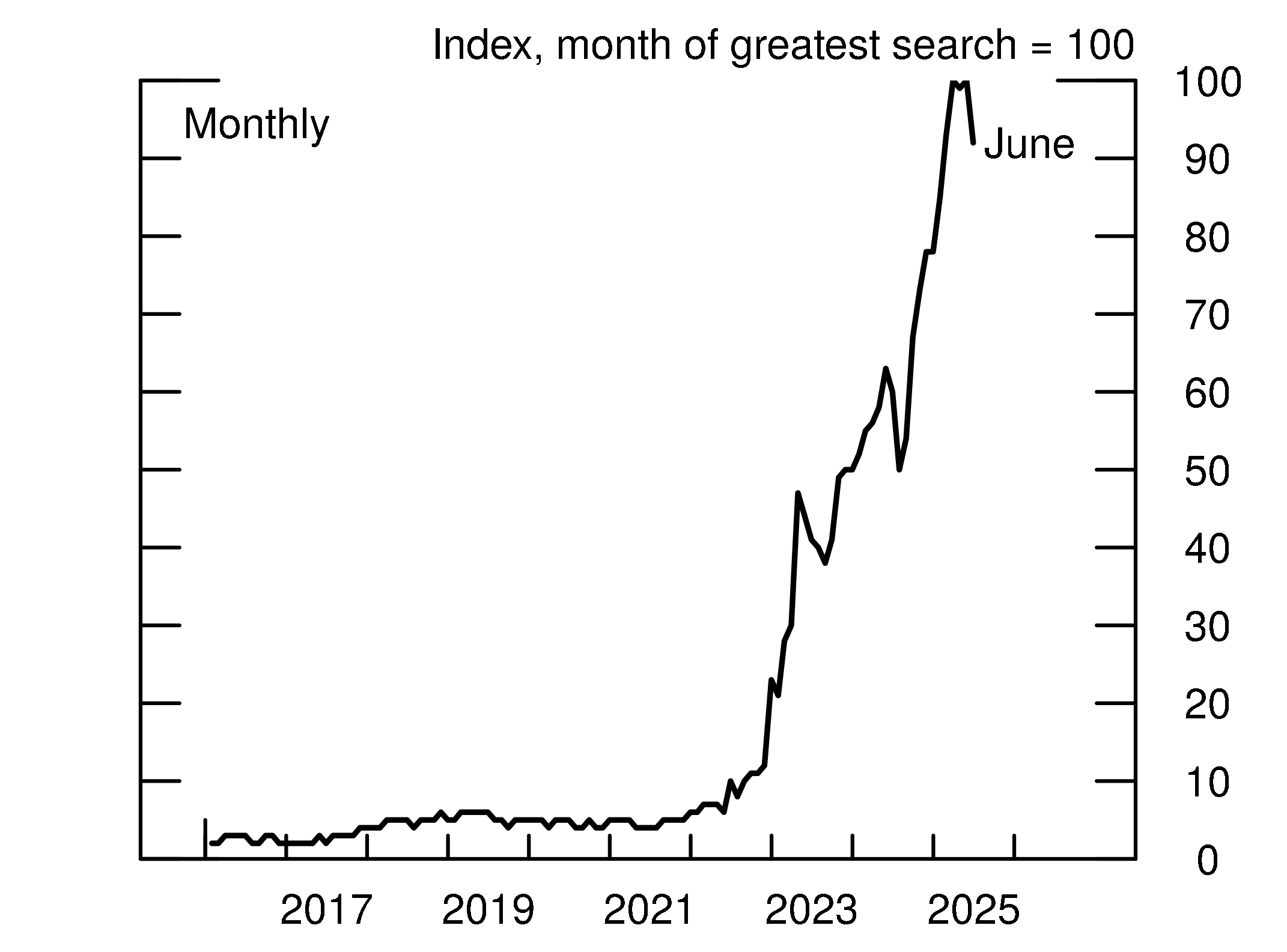}
\end{subfigure} 
\begin{minipage}{\linewidth}\setlength{\belowdisplayskip}{0pt} \setlength{\abovedisplayskip}{0pt}
\caption*{\scriptsize{Note: Apps are ChatGPT, Claude, DeepSeek, and Perplexity.  Includes Android and iOS.  Android download information not available for China.  Does not account for access via application program interface.  Web searches include related terms in Google's ``AI'' topic.\\
Source:  appfigures; Google Trends.}}
\end{minipage}
\end{figure}

\begin{figure}
\centering
\caption{\label{figure:ai-related-investment}Indicators of U.S. AI-Related Investment}
\begin{subfigure}[t]{0.48\textwidth}
\caption{\label{figure:data-center-construction}Data Center Construction}  
\includegraphics[width=\textwidth]{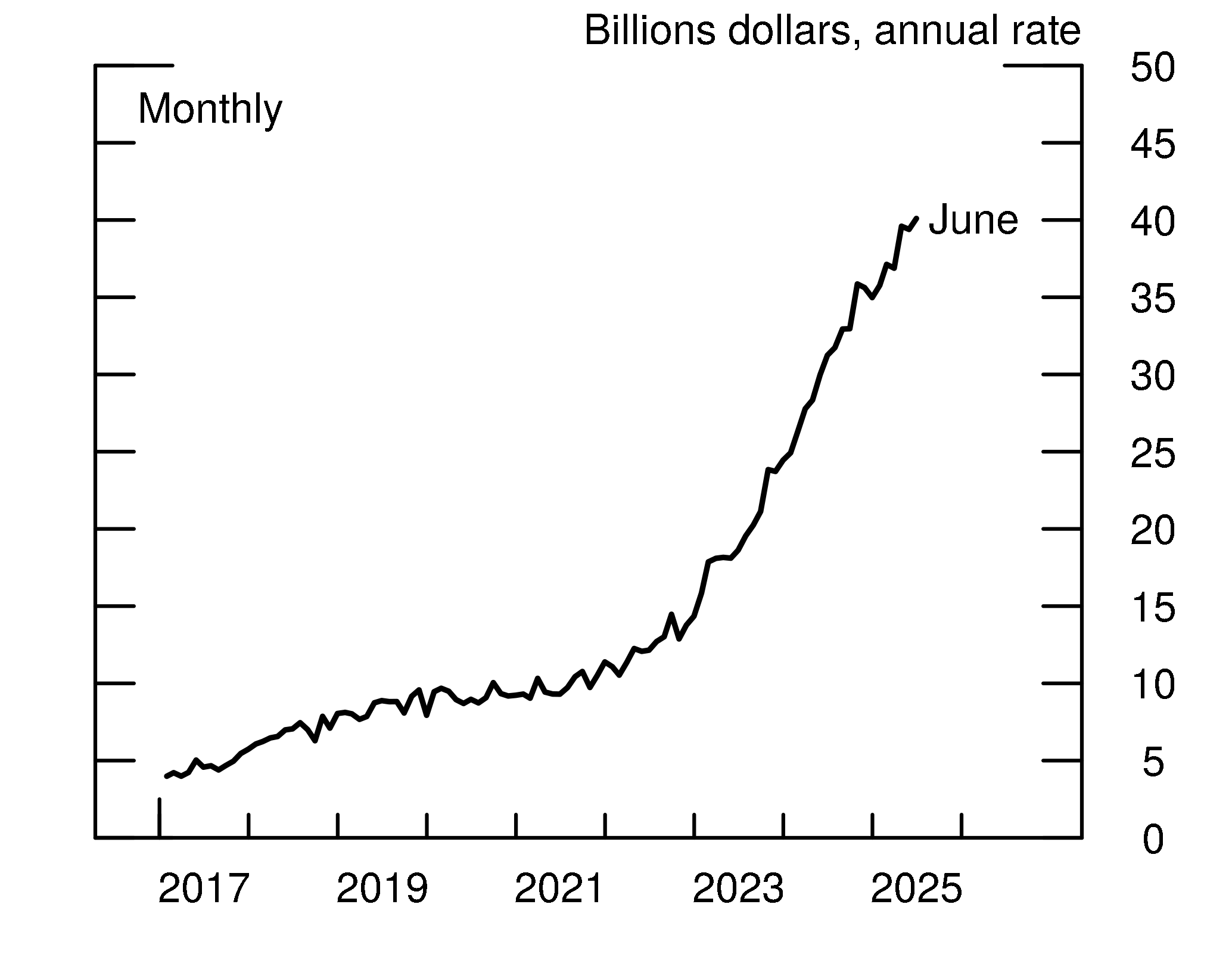}
\end{subfigure} 
\begin{subfigure}[t]{0.48\textwidth}
\caption{\label{figure:chip-absorption-us-asic}Application-Specific Chips}
\includegraphics[width=\textwidth]{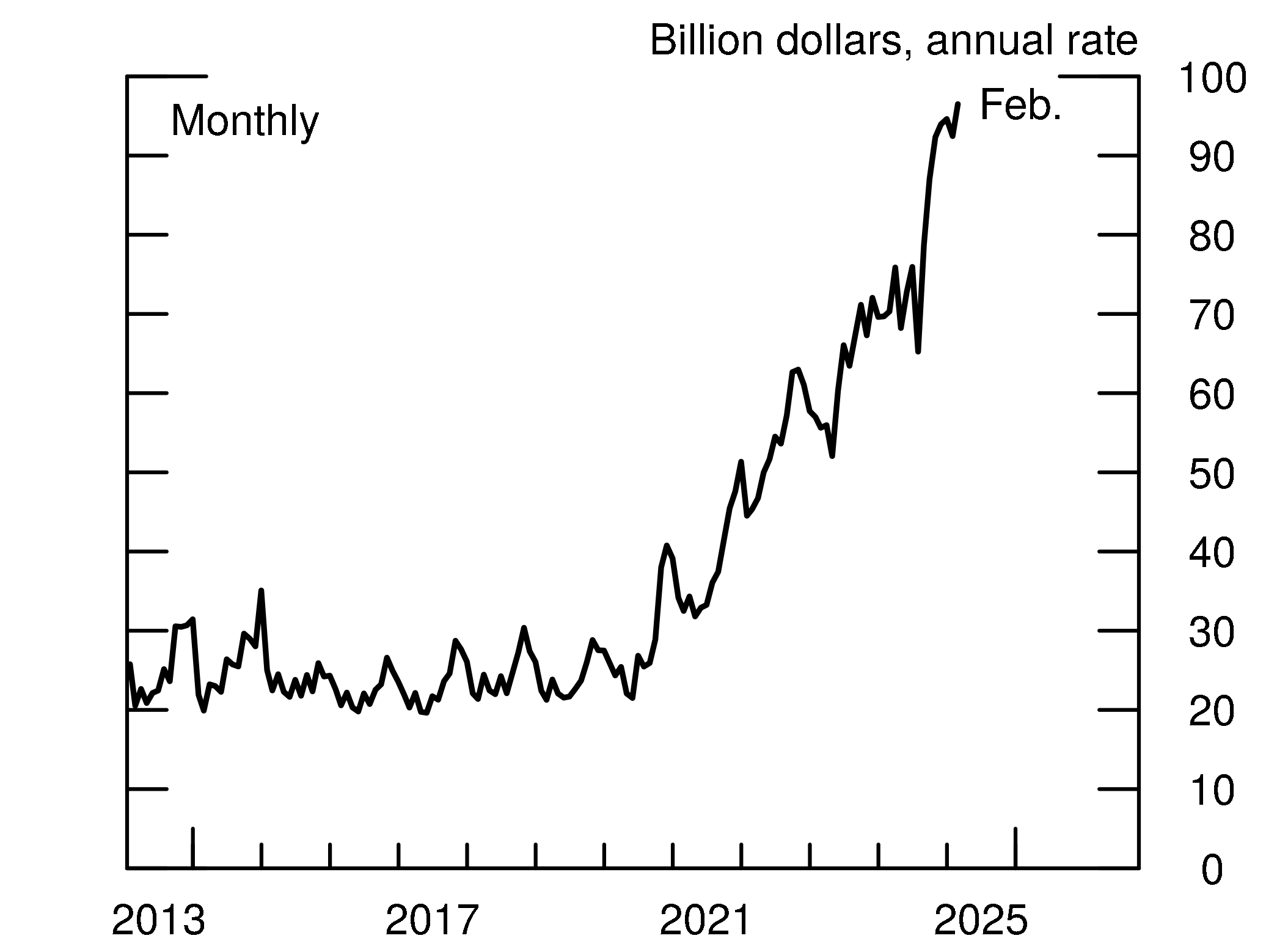}
\end{subfigure}
\begin{minipage}{\linewidth}\setlength{\belowdisplayskip}{0pt} \setlength{\abovedisplayskip}{0pt}

\caption*{\scriptsize{Note:  Nominal value of construction put in place.   Application-specific chips include GPUs, TPUs, and ASICs for other applications.\\Source:  Census Bureau; Semiconductor Industry Association.}}
\end{minipage}
\end{figure}

Web searches for AI and downloads of the ChatGPT app have soared, sparking intense competition for leadership in genAI (\vref{figure:indicators-interest-genai}), and the computational intensity of training genAI models and processing user requests has led to a massive increase in data center construction and spending on AI-related semiconductor chips (\vref{figure:ai-related-investment}). While optimists see potential for genAI to spur an information technology (IT)-fueled productivity boom comparable to the late 1990s and early 2000s, more downbeat observers see claims about its capabilities as overstated and highlight potential headwinds, such as regulations to guard against unintended harms, political pushback as AI affects the jobs of human workers, and the colossal energy requirements for training and running genAI systems.\footnote{For an optimistic view, see \citet{kurzweil:2024:singularity-nearer-merge-ai}. For a more cautious perspective see \citet{narayanan-kapoor:2024:ai-snake-oil-artificial-intelligence}.}  It is too early to adopt either view with confidence.  Whether practical genAI applications will be consequential enough to raise aggregate productivity growth remains to be seen.  GenAI may be no more important than previous innovations in IT already reflected in the historical trend, including its predecessors in the field of AI.\footnote{Moreover, the present era of modest productivity growth in the midst of mature machine learning, cloud computing, and smartphones should temper expectations for another IT boom.}

With only ``green shoots'' of quantitative evidence in hand that genAI will raise productivity, we frame the prospective effect of genAI in qualitative terms: We ask what class of innovation it may be.  Some labor-saving innovations, such as the light bulb, temporarily raise productivity growth as adoption spreads, but the effect fades when the market is saturated; that is, the level of output per hour is permanently higher but the growth rate is not.  Other widely used technologies, such as the electric dynamo, spur knock-on innovations---new products, process improvements, and business reorganization---and refresh this adoption cycle through ongoing improvement in the core technology \citep{david:1990:dynamo-computer-historical-perspective-modern}.  The boost to productivity growth from these general-purpose technologies (GPTs) may last longer.  Yet other inventions, such as the compound microscope, increase the efficiency of the research and development process; these ``inventions of methods of invention'' (IMIs) yield a sustained increase in productivity growth by lowering the cost of research and development.

We first define ``generative AI,'' then consider the evidence that genAI is a GPT, reviewing indicators for the scope of diffusion, the extent of knock-on innovations, and signs of ongoing progress in the core technology.  To assess its status as an IMI, we discuss evidence that it increases the efficiency of observation, analysis, communication, and organization in research and review several indicators (patents, earnings calls, and query topics). For both questions, we reference case studies for the financial, health care, and information sectors and for the electricity generation industry \citep{baily-kane:2025:ai-finance-sector-transforming-productivity, baily-kane:2025:ai-healthcare-sector-enhancing-care, kane-baily:2025:ai-electricity-sector-optimizing-grid, kane-baily:2025:ai-information-sector-advancing-software}.   We conclude there is substantial evidence that genAI is both a GPT and an IMI, an encouraging sign its adoption will lead to higher productivity in the future.  

There is a substantial literature on the question of whether machine learning, which preceded genAI, may be a GPT \citep{cockburn-henderson-stern:2018:impact-artificial-intelligence-innovation-exploratory,trajtenberg:2018:ai-next-gpt-political-economy,bresnahan:2019:artificial-intelligence-aggregate-growth-prospects,goldfarb-taska-teodoridis:2023:could-machine-learning-general-purpose,bresnahan:2024:innovation-paths-ai-gpt}, and \citet{cockburn-henderson-stern:2018:impact-artificial-intelligence-innovation-exploratory} discuss the possibility that machine learning is an IMI.  There is little work focused specifically on genAI.  \citet{eloundou-manning-mishkin-rock:2024:gpts-gpts-labor-market-impact}, a prominent exception, consider the prospects for genAI to be a GPT; relative to that work, we draw on a broader set of indicators and consider evidence for whether genAI is both a GPT and an IMI.   Our focus on characteristics of the technology itself and its integration into business processes complements the large literature on the labor market impact of genAI.\footnote{On labor market effects of AI, see \citet{acemoglu-restrepo:2020:wrong-kind-ai-artificial-intelligence,agrawal-gans-goldfarb:2023:turing-transformation-artificial-intelligence-intelligence,brynjolfsson-li-raymond:2025:generative-ai-work,eloundou-manning-mishkin-rock:2024:gpts-gpts-labor-market-impact}, among others.}  Our qualitative assessment of genAI also serves as a primer to inform discussion of AI in the context other literatures, such as creative destruction and endogenous growth \citep{akcigit-van-reenan:2023:economics-creative-destruction}.\footnote{Our focus is primarily on the U.S. economy; \citet{filippucci-etal:2024:impact-artificial-intelligence-productivity-distribution} take a global perspective and address many of the same issues as our paper.}

\section{What is Generative AI?}\label{section:what-is-generative-ai}

``Artificial intelligence" (AI) is an umbrella term encompassing a variety of algorithms deployed on computers to mimic human thought, communication, and choices, such as machine learning, computer vision, and generative models.  (See Appendix \ref{appendix:definitions-ai} for a discussion of several influential definitions of AI.)  AI systems achieve these objectives by constructing and calibrating mathematical models of complex patterns found in training data.  The most widely known implementations of genAI use a computer model of the human discourse found on the internet to respond to natural-language prompts (questions or directives), though genAI systems take other forms as well as we discuss below.

We narrow our focus to genAI for several reasons.  First, the broad and varied use of ``AI'' makes a coherent discussion of its effects on productivity difficult.  Second, the productivity impact of genAI is largely in the future, in contrast to AI types already in use for which the effects on productivity are, in principle, an empirical question.  Third, the human-like behavior of generative models has made concerns about the disruptive effects of AI particularly salient.

Although systems which respond to prompts with natural language text were a part of the AI field from its inception, early models were not grounded in a model of human language. (For a short history of the field, see Appendix \ref{appendix:history-ai}.) For example, rudimentary chatbots, such as ELIZA the psychotherapist, text autocompletion, and ``expert systems,'' such as MYCIN for infection diagnosis, appeared in the 1960s and 1970s.  These systems had matured by the 2010s, when sophisticated voice-driven chatbots like Alexa and Siri had been introduced, news outlets were auto-generating routine stories, and IBM's Watson, famous for beating humans at \textit{Jeopardy} in 2011, was repurposed to provide advice on a host of topics.  These were symbolic, rules-based systems, albeit with an element of randomness in their output.

Richer generative models appeared after the development of large language models (LLMs). LLMs, which represent the meanings of words and their relationships by locations in a high-dimensional space, emerged in the 2010s.  Word2Vec, most notably, encoded words as vectors of numerical values, and while the values do not correspond to specific, interpretable characteristics, they capture semantic relationships in an abstract space \citep{mikolov-chen-corrado-dean:2013:efficient-estimation-word-representations-vector}.  For example, while humans would represent a dress by its color, size, and hem length, say, the characteristics chosen by Word2Vec would not have a readily apparent interpretation \citep{bajari-chernozhukov:quality-adjusted-price-indexes-ml:2018}.  Aided by advances in computational power and big data processing techniques, genAI developers pushed the field forward and achieved a breakthrough with the Transformer architecture in 2017 \citep{vaswani-shazeer-parmar:2017:attention-all-you-need}.  (The technical features of the Transformer are discussed in a box below.)  This model allows for a richer representation of word meaning in its encoding by accounting for context.  Many GenAI models use LLMs to encode the tokens (words, phrases, or parts of words) in the input in this fashion.  After embedding the input text into locations within a high-dimensional space of abstract characteristics, they draw on the information embedded nearby to guide the prediction of the next token, weaving the output into a relevant natural-language response. 

Because these models are created as neural networks, which are extremely flexible, they differ from previous models, like IBM's Watson, in that they are not symbolic. That is, they do not have a predetermined logical structure.  The effect of this added flexibility and the richness enabled by their massive scale is that the range of possible generated content is more open-ended than earlier systems.  For example, earlier symbolic attempts were capable of producing formulaic news stories about a firm's quarterly earnings or the outcome of a baseball game, but ChatGPT can provide a convincing nuanced response to prompts such as ``write a short story about an angst-ridden robot in the style of Edgar Allan Poe.''

Importantly, genAI can support a wider array of applications than natural language tools for learning and creativity.  Broadly stated, genAI systems produce contextually appropriate artifacts using an open-ended stochastic process that draws from patterns in a dataset.  The artifacts may be a variety of things other than text, including computer code, images, music, chemical structures, game environments, mathematical proofs, or dance moves, to name a few.  And, while the chat window user interface makes genAI accessible, it is not found in all applications.  For example, in generative adversarial networks (GANs), neural networks interact with each other, not humans, to generate output.

\begin{tcolorbox}[every float=\centering, breakable,enhanced,title = Landmark AI Models: The Transformer, title after break = The Transformer (continued), parbox = false]\label{box:transformer-ai-model-architecture}
The transformer architecture, introduced by  \citet{vaswani-shazeer-parmar:2017:attention-all-you-need}, was a game changer in AI, particularly as the engine behind genAI models. Its key innovation, the ``attention mechanism,'' steers models to focus selectively on relevant parts of the prompt, enabling more efficient and accurate processing of language. This breakthrough has powered major advancements in natural language understanding, translation, and generation, forming the backbone of today's most advanced genAI systems.

Transformers process input data through a series of layers (steps), each consisting of an attention mechanism followed by a multilayer perceptron (MLP, defined below), proceeding as follows.  

First, a representation of the prompt (input text) suitable for analysis by the model is created. Specifically, the prompt is broken into tokens (smaller pieces which may be phrases, words, or parts of words).  The tokens are converted into embeddings (numerical vector representations) which encode the semantic and syntactic meaning of each token. Loosely speaking, for each token, the closest of the other tokens, as measured by the distance between their embeddings, are the ones most important to understanding its meaning.

Second, the attention mechanism processes the matrix of token embeddings using three large matrices called the “query,” the “key,” and the “value.” For each token in the input, the query is compared to the keys of all tokens to compute attention scores, which are used to form a weighted average of values. This step allows each token's representation to incorporate information from other tokens in the prompt based on their contextual relevance. 

Third, the data passes through an MLP, a type of neural network. While the attention mechanism focuses on pairwise interactions between tokens, the MLP applies nonlinear functions (in contrast to the linear attention mechanism) in refining the token representations. 

This sequence---of computing the attention mechanism followed by the MLP---is repeated multiple times depending on how many layers are in the model (for example, the Llama-3 model has 32 layers), enabling the model to capture increasingly abstract features of the input text. 

The performance gains from scaling of this system through increasing the size of these matrices---along with larger training datasets and improvements in hardware and processing algorithms---underpins the rising ability to handle complex language tasks.
\end{tcolorbox}

\section{Is GenAI a General Purpose Technology?}

While the evidence for genAI-driven productivity in specific tasks is intriguing (\vref{table:selected-genai-studies}), to look ahead to its future productivity impact, we would like to know (1) if genAI will be widely adopted, (2) the extent of related innovations, and (3) whether genAI will continue to improve.  That is, we would like to know if genAI is a general-purpose technology (GPT).\footnote{Note that the ``GPT'' initialism in the names of OpenAI genAI models stands for ``generative pre-trained transformer.''  We will use ``GPT'' to mean ``general-purpose technology'' exclusively in this paper except when referring to OpenAI models.}  Through  widespread adoption, downstream complementary innovation, and sustained innovation in the core technology, GPTs have long-lasting effects on productivity.  Examples of GPTs are shown in \vref{table:examples_of_gpts}.  

As described by \citet[xvi]{lipsey-carlaw-bekar:2005:economic-transformations-general-purpose-technologies}, ``big GPT shocks change almost everything in a society and revitalize the growth process by creating an agenda for the creation of new products, new processes, and new organizational forms.''  For example, new products that followed the development of the (electronic) computer include office productivity software, ATM machines, and the routing equipment that directs traffic around the internet.  New processes that followed the development of reliable electricity include the production of hydrogen by electrolysis, salvaging scrap steel using electric arc furnaces, and the fabrication process for semiconductor chips.  And, a new organizational form that followed the introduction of the three-masted sailing ship was the joint-stock company, used to finance the voyages of large-scale trading firms. 

We briefly describe the three criteria for a technology to be a GPT, then examine in detail the evidence that genAI meets each one.

\begin{table}[t]
    \centering
    \caption{Examples of General Purpose Technologies}\label{table:examples_of_gpts}
    \begin{tabular}{|l|l|}
    \hline
     Technology & Initial Impact \\
     \hline
     Domestication of plants    & 9000--8000 BCE \\
     Writing    & 3400--3200 BCE\\
     Iron    & 1200 BCE \\
     Waterwheel    & Early medieval period\\
     Three-masted sailing ship    & \nth{15} century\\
     Printing    & \nth{15} century \\
     Factory system & Mid \nth{18} century \\
     Steam engine & Late \nth{18} century\\
     Railway & Mid \nth{19} century \\
     Internal combustion engine & Late \nth{19} century \\
     Electricity & Early \nth{20} century \\
     Motor vehicle & Early \nth{20} century \\
     Mass production, continuous process factory & Early \nth{20} century \\
     Lean production & Late \nth{20} century \\
     Computer & Late \nth{20} century \\
     Internet & Late \nth{20} century \\
    \hline
    \multicolumn{2}{|l|}{Source: Adapted from \citet{lipsey-carlaw-bekar:2005:economic-transformations-general-purpose-technologies}.} \\
    \hline
    \end{tabular} 
\end{table}

\paragraph{Diffusion} The more widespread the application of a technology, the greater the potential impact on aggregate productivity \citep{hulten:1978:growth-accounting-intermediate-inputs}.  That said, however widely adopted, the \textit{direct} productivity effect of a single invention is bounded.  Productivity \textit{growth} will be higher during the adoption transition, but will return to its underlying trend when diffusion is complete \citep{solow:contribution-theory-economic-growth:1956}.  To illustrate using the light bulb, once suitable filaments were developed and the inexpensive incandescent light bulb was available, the technology was gradually adopted in workplaces, raising productivity through better visibility and lower risk of accident \citep{abdou:1997:effects-luminous-environment-worker-productivity}.  Lighting is necessary for nearly all human labor, so the potential effect on the level of productivity from the light bulb was noteworthy, but once the light bulb market was saturated, it delivered no further (direct) level effect and the increment to productivity growth present during the transition disappeared.\footnote{The light bulb was not solely a terminal innovation, of course.  The surge in demand for electricity represented by light bulbs led to centralized power stations \citep{david:1990:dynamo-computer-historical-perspective-modern}.}

\paragraph{Knock-on Innovation} 
Technologies that spur further innovation can deliver a longer-lived impetus to productivity growth.  The greater persistence of elevated growth is the result of a series of overlapping classical ``light bulb'' growth effects.\footnote{Complementary innovations may increase productivity by raising the effectiveness of the GPT as well, such as raising the operating rate of computers by the invention of cloud computing.}  The electric dynamo is an example.  The dynamo uses electromagnetism to convert mechanical energy produced by a prime mover---a steam engine, say---to electromagnetic energy, which is then conveyed by wires and converted back to mechanical energy by a motor used to drive machinery in another location.  Existing systems conveyed mechanical energy directly to machinery through a set of belts.  The dynamo/wiring/motor system is more energy efficient than the belt system except in very simple arrangements, so the simple replacement of existing factory systems yielded productivity gains.  In addition, the dynamo enabled a more flexible organization of production \citep{david:1990:dynamo-computer-historical-perspective-modern}.  The less centralized factory designs adopted by firms in response are a knock-on productivity-enhancing innovation spurred by the dynamo.

\paragraph{Ongoing Core Innovation} When a technology continues to improve over time, the new target productivity level---fixed in the simple impulse-response framework of the \citet{solow:contribution-theory-economic-growth:1956} model---becomes a moving target.  Ongoing innovation translates into greater technical performance at a lower cost, a form of productivity gain.  Moreover, the price of capital typically follows the production cost downward spurring greater adoption.  Innovation is, in a sense, embedded in the capital \citep{kaldor:1957:model-economic-growth}.  Solid state electronics is an example.  Relentless increases in the number of transistors on each semiconductor chip has driven the price of computing lower, making it cost-effective both to embed electronics in a greater variety of devices (like inexpensive toys) and to enhance devices with more and more electronic capability (like smartphones). 

\subsection{Diffusion}

Although comprehensive measures of the diffusion of genAI, specifically, are limited, recent trends in the diffusion of AI, generally, may indicate the underlying influence of genAI.  Surveys show AI adoption rising, particularly in large corporations where AI use is concentrated.  Even so, a large majority of firms still don't see an application for AI in their business.  Analyses of the text of job descriptions suggest that AI can be used for a broad range of workplace tasks, indicating that the potential for diffusion among firms is high.\footnote{See \citet{acemoglu-autor-hazell-restrepo:2020:ai-jobs-evidence-online-vacancies}, \citet{brynjolfsson-mitchell-rock:2018:what-can-machines-learn}, \citet{felten-raj-seamans:2019:occupational-impact-artificial-intelligence-labor}, \citet{webb:2019:impact-artificial-intelligence-labor-market}, \citet{eloundou-manning-mishkin-rock:2024:gpts-gpts-labor-market-impact}.}  At the same time, the share of job postings mentioning AI skills is modest, indicating that firms are taking a cautious approach to hiring workers to focus on AI use.  Meanwhile, from the worker's perspective, AI adoption seems widespread; surveys of individuals document that a large share of workers are already AI users.    

\begin{figure}[ht]\caption{AI Use Over Time}\label{figure:ai-adoption-over-time}
\centering
\begin{subfigure}[t]{0.48\textwidth}
\caption{\label{figure:ai-adoption-btos}Census BTOS (United States)}   
\includegraphics[width=\textwidth]{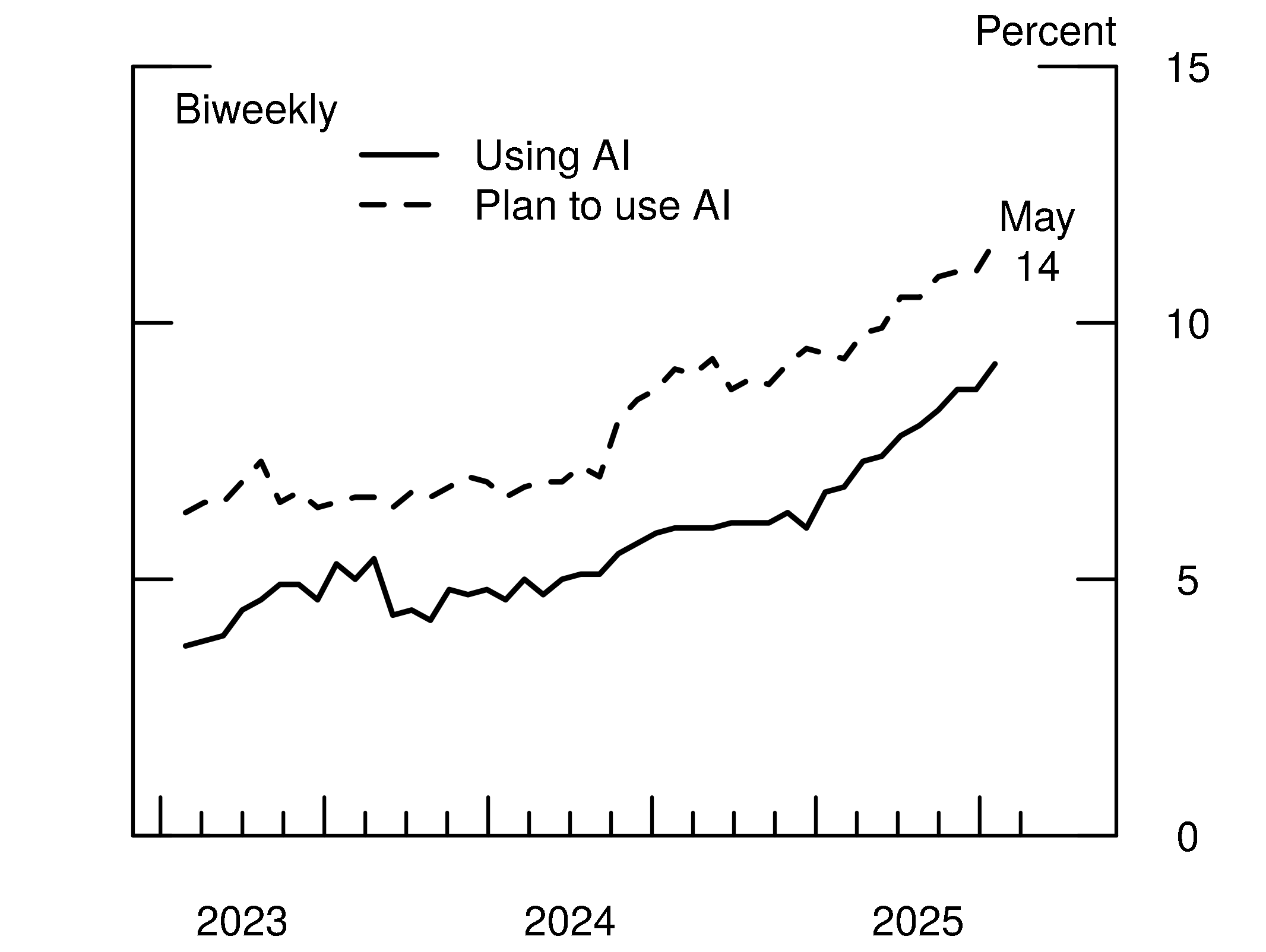}
\end{subfigure}
\begin{subfigure}[t]{0.48\textwidth}
\caption{\label{figure:ai-use-mckinsey}McKinsey (Global)}   
\includegraphics[width=\textwidth]{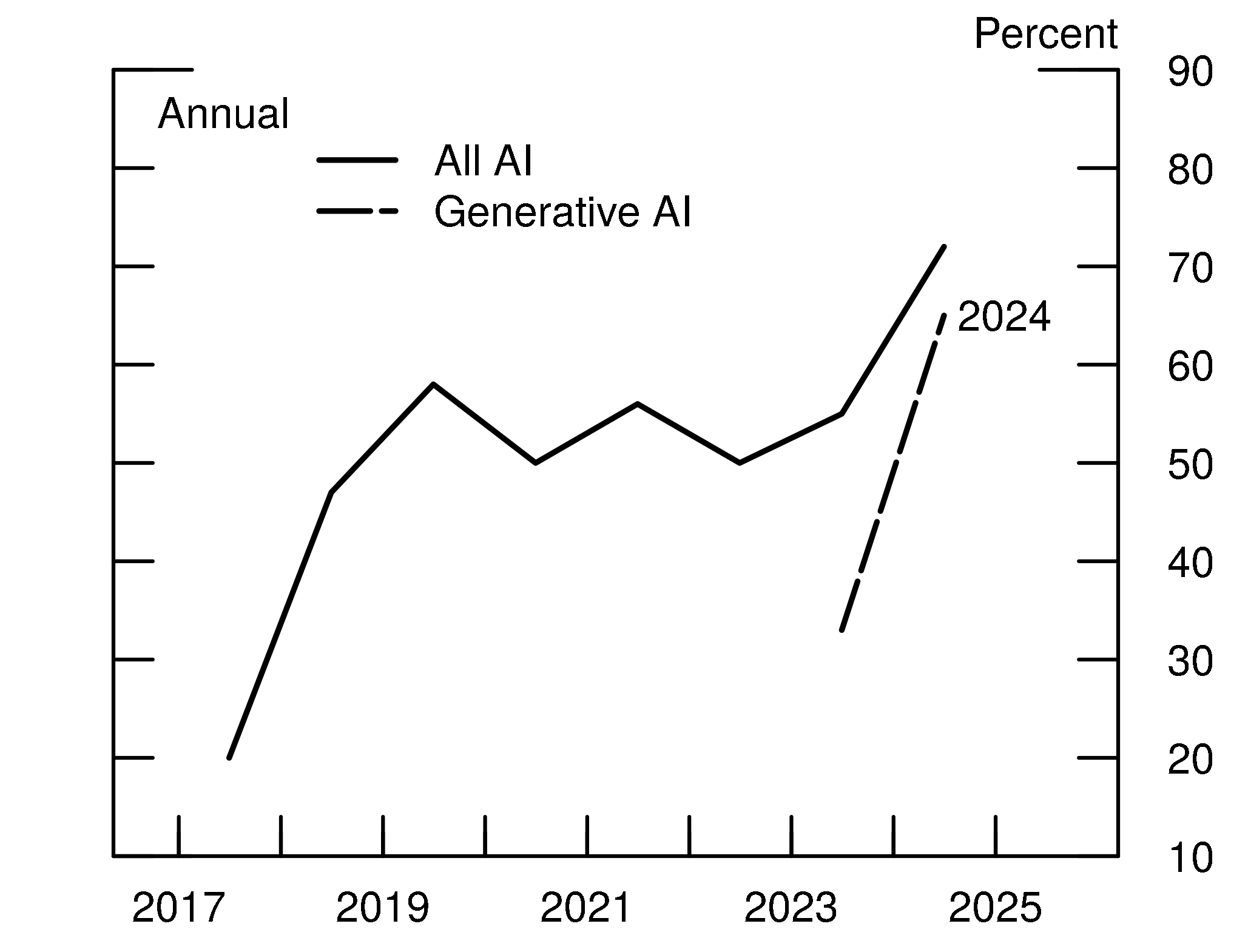}
\end{subfigure}
\begin{minipage}{\linewidth}\setlength{\belowdisplayskip}{0pt} \setlength{\abovedisplayskip}{0pt}
\scriptsize{\vspace{2em}Note: For BTOS, respondents were asked about AI use in producing goods or services during the past two weeks and anticipated in the next six months.  For McKinsey, respondents were asked if they ``use AI in at least one business function''\\Source: Census Bureau, Business Trends and Outlook Survey; McKinsey, ``The State of AI in Early 2024.''}
\end{minipage}
\end{figure}

\paragraph{Adoption surveys}  Distilling a single message from the available surveys of AI use is difficult at first glance. The Census Bureau's Business Trends and Outlook Survey (BTOS) finds roughly 9\% of firms use AI, while McKinsey reports that 72\% of firms do so (\vref{figure:ai-adoption-over-time}).  On closer inspection, these surveys are consistent with one another and reveal important nuances in the state of AI adoption with respect to firm size and business functions.

Combining the results of these surveys points to far higher AI adoption for large firms than small ones.\footnote{There may well be significant heterogeneity within small firms on this question.  Because new firms, which are typically small, may begin life as digitally native firms, they may adopt AI more easily.  The pace of business applications with a high-propensity of turning into businesses with payroll, reported by the Census Bureau, has moved up significantly in the wake of the pandemic.  Note that information on new firms will appear with a delay in the BTOS, for which the sample is drawn from the Census Business Register \citep{bayard-etal:2018:early-stage-business-formation-analysis}.}  The BTOS is a representative sample of 200,000 U.S. firms, only a handful of which are large corporations \citep{bonney-etal:2024:tracking-firm-ai-real-time}. The McKinsey survey, in contrast, is a convenience sample with heavy representation from large corporations \citep{mckinsey:2024:state-ai-early-2024}.\footnote{Although McKinsey states that its survey includes ``participants representing the full range of regions, industries, company sizes, functional specialties, and tenures,'' responses are not weighted by the relative prevalence of their characteristics in the population.  We thank Michael Chui for confirming that this is a fair characterization of the survey.}  The U.S. firm size distribution is highly skewed and large corporations have thousands of employees \citep{kondo-lewis-stella:2024:heavy-tailed-zipf-firm-establishment}.  The threshold for the largest firm-size group in the BTOS is 250 employees, for which BTOS reports 14.9\% adoption, and the BTOS reports that for smallest firm-size group, firms with fewer than 5 employees, 9.7\% were using AI in June 2025.

These surveys also suggest that AI use may be less prevalent in core business functions than in support functions.  Differences in the definition of adoption between the surveys point to this conclusion.  The BTOS survey asks, ``In the last two weeks, did this business use Artificial Intelligence (AI) in producing goods or services? (Examples of AI: machine learning, natural language processing, virtual agents, voice recognition, etc.).'' The McKinsey question is rather less restrictive, asking respondents if they use ``AI in at least one business function.'' That is, the BTOS asks about use in core business functions (``producing goods or services''), while McKinsey includes non-core functions like ``marketing and sales,'' which is the function with greatest AI use in their survey.  

Evidence about adoption for genAI specifically is more limited.  Only the McKinsey survey asks separately about genAI, finding that use among their (primarily large firm) respondents surged from roughly one-third in 2023 to two-thirds in 2024. \citet{bick-blandin-deming:2024:rapid-adoption-generative-ai} survey workers (a representative sample of 18-64 year-olds in the United States) and find nearly 40\% of respondents used genAI.

As posited in \citet{crane-green-soto:2025:measuring-ai-uptake-workplace}, high-level managers may underestimate the extent to which their employees are using AI tools, or AI users may be highly concentrated in large firms (roughly half of the U.S. workforce) or AI-intensive industries.  Consistent with the idea that AI users may be concentrated in AI-intensive industries, \citet{sergeyuk-golubev-bryksin-ahmed:2025:using-ai-based-coding-assistants} survey programmers and find 84\% of them use AI.  A recent trend in industry composition is suggestive as well.  \citet{decker-haltiwanger:2024:high-tech-business-entry-pandemic} identify a step up in the share of business establishments---reported by the Bureau of Labor Statistics in the \textit{Quarterly Census of Employment and Wages}---in industries with a high share of employees in science, technology, engineering, and math fields.  We speculate this employee composition may be paired with higher AI use.

\paragraph{Job postings}

\begin{figure}[ht]\caption{AI-Related Job Postings}\label{figure:ai-related-job-postings}
\centering
\includegraphics[width=0.5\textwidth]{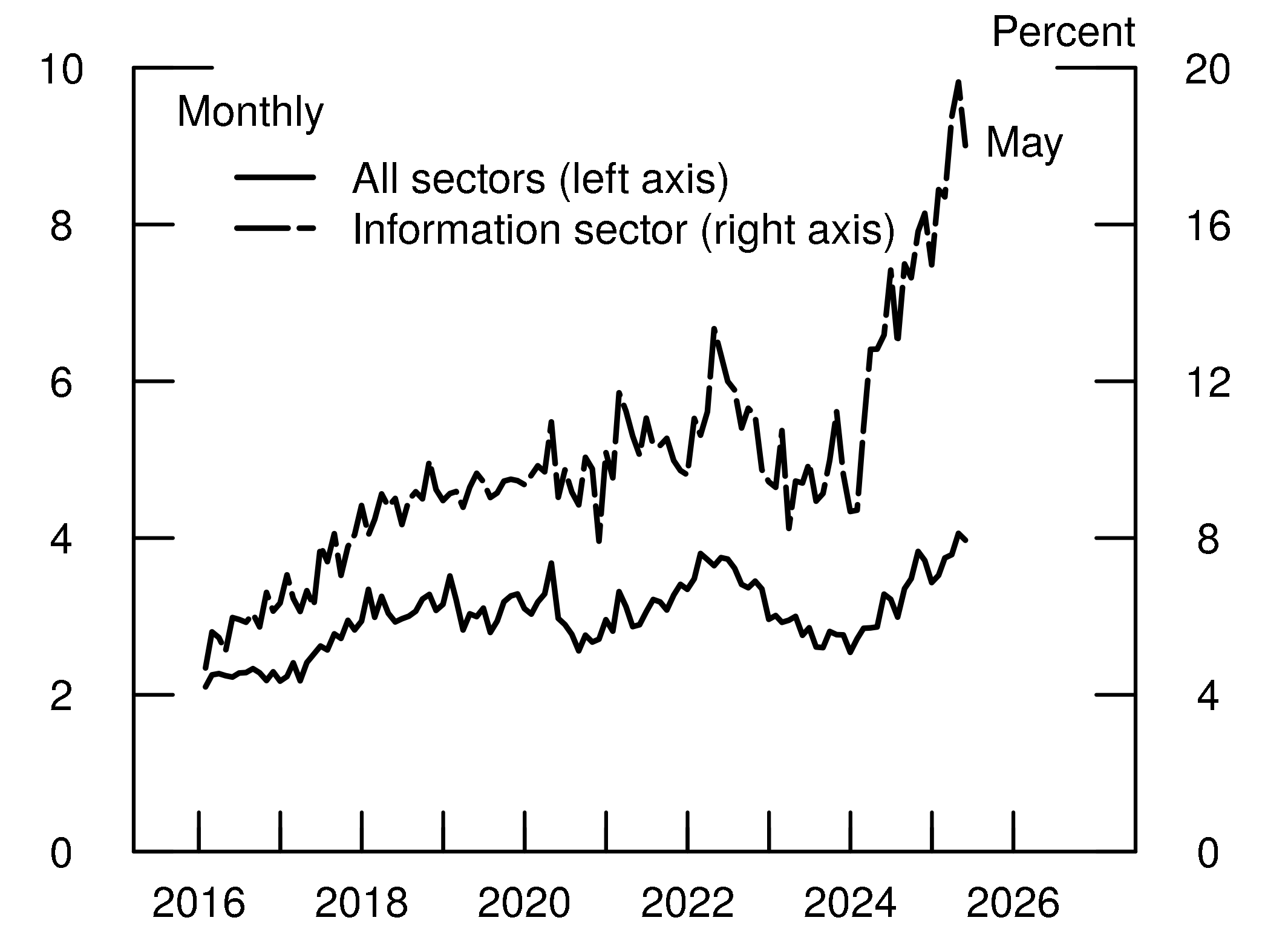}
\begin{minipage}{\linewidth}\setlength{\belowdisplayskip}{0pt} \setlength{\abovedisplayskip}{0pt}
\scriptsize
		{\vspace{2em}Note: Jobs classified using AI-related terms found in job descriptions, as described in \citet{acemoglu-autor-hazell-restrepo:2020:ai-jobs-evidence-online-vacancies}. List of AI-related terms updated by Lightcast. \\Source: Lightcast.}
\end{minipage}
\end{figure}

Job posting data from Lightcast (formerly Burning Glass) categorized using terms associated with AI by \citet{acemoglu-autor-hazell-restrepo:2020:ai-jobs-evidence-online-vacancies}, extended with terms not prevalent at the time of their analysis (such as ``genAI''), show the share of job postings currently related to AI to be  roughly 4 percent, using a broad definition of AI including a cluster of related skills (\vref{figure:ai-related-job-postings}).  Importantly, that share moved up (from no more than 2 percent) around 2017, well before practical genAI applications became prevalent, consistent with earlier forms of AI driving the increase.  AI-related job postings have moved up only modestly since then, suggesting that explicit use of genAI-related skills is not crucial to many jobs.  In some sectors, most notably the information sector (shown in the graph), the share of job postings referencing AI is substantially higher.

\paragraph{Case study evidence}\textcolor{white}{x}

\vspace{2mm}

The \textbf{information sector} has adopted genAI rapidly. Among U.S. programmers, 92\% were using AI tools (including genAI) as of June, 2023 \citep{shani-github:2023:survey-reveals-ais-impact-developer}. Other occupations within the information sector use genAI in their work as well. Of graphic designers and illustrators, 69\% used genAI in their work in 2023, employing tools like generative fill and large-scale text-to-image models such as DALL-E.\footnote{See Offerman, Stefan. ``Creative Pros See Generative AI as Part of Their Future.'' Adobe Blog, March 21, 2023. \url{https://blog.adobe.com/en/publish/2023/03/21/research-creative-pros-see-generative-ai-as-part-of-their-future}.} 

In the U.S. \textbf{health care} system, adoption has been slow for IT generally and AI in particular \citep{poon-etal:2006:assessing-level-healthcare-information-technology,goldfarb-taska-teodoris:2020:artificial-intelligence-health-care-evidence}. However, genAI may be an exception:  A majority of physicians already used or were planning to use genAI in 2025 for generating chart summaries, creating discharge instructions, and an array of other tasks \citep{AMA:2025:physician-sentiments-ai-health-care}.  Radiology is an enlightening example of how genAI has changed AI use in health care.  While approximately 30\% of radiologists already used AI as of 2020 \citep{allen-agrawal-coombs-wald-dreyer:2020:acr-data-science-institute-artificial}, the emergence of genAI spread AI to all stages of the radiology workflow, including the health system (e.g. imaging need prediction, claims processing), clinicians (e.g. prior authorization, communication), technologists (e.g. patient tracking, report generation), and radiologists (e.g. image quality assessment, diagnosis) \citep{burnside-grist-lasarev-garrett-morriss:2025:artificial-intelligence-radiology-leadership-survey}.

Notwithstanding these inroads, AI use faces substantial hurdles in health care.  For example, \citet{agarwal-moehring-rajpurkar-salz:2023:combining-human-expertise-artificial-intelligence} document that combining human analysis with AI diagnostics can yield disappointing results---radiologists have difficulty determining how much confidence to assign the AI output.  And, although \citet{sahni-stein-zemmel-cutler:2023:potential-impact-artificial-intelligence-healthcare} conclude that machine learning has the potential to assist with an array of tasks including diagnosis, treatment choice, and managing records, they assess that for hospitals, AI adoption often cannot be justified on financial factors alone.

GenAI is used for many tasks in \textbf{finance} as well.  Companies can use genAI to lower the cost of creating client-specific portfolios \citep{joshi:2025:generative-ai-investment-portfolio-management}, improve existing automated systems that respond to client requests \citep{mckinsey:2025:banking-innovation-ing-generative-ai}, assist with regulatory compliance \citep{agarwal-etal:2024:generative-ai-banks-manage-risk} and with loan underwriting \citep{wang:2023:banks-credit-unions-testing-ai}.

In \textbf{electricity generation and distribution}, firms are experimenting with genAI, with approximately 33\% piloting genAI in their customer service operations.\footnote{Penrod, Emma. ``A Third of Utilities Have Begun to Pilot Generative AI for Customer Service, Other Uses: Report.'' Utility Dive, July 12, 2023. \url{https://www.utilitydive.com/news/utilities-generative-ai-artificial-intelligence-capgemini-report/686601/}.} Some companies use genAI in their load forecasting processes, while others use the technology to simulate equipment degradation and inform predictive maintenance, areas where machine learning was already in use \citep{gao-etal:2024:general-framework-load-forecasting-pretrained}.

\subsection{Knock-on Innovation}

In the course of adopting new technologies, firms retool, retrain, and reorganize to better exploit their productivity potential.\footnote{\citet{bresnahan-greenstein:1996:technical-progress-coinvention-computing-computers} discuss this process in detail, which they refer to as ``co-invention.''}  This process involves knock-on innovation in products, production processes, and operations.  In the case of genAI, knock-on product innovation includes the diverse array of user interface software for generative models and the emergence of more capable robots. Process innovation includes genAI-enhanced approaches to product design and production line operation.  Organizational innovations include centralized data governance and optimization of supply chains and data centers.  

\citet{bresnahan:2024:innovation-paths-ai-gpt} observed that early machine learning (pre-generative AI) adoption was concentrated in places where complementary innovation was less necessary, such as in firms that were highly digitized from their founding (digital natives).  In such firms, adoption of AI was more straightforward, involving substitution of AI for existing IT capital or deploying AI to undertake tasks not previously part of operations.\footnote{The legendary email from Jeff Bezos instructing internal teams at Amazon to exclusively use APIs to deliver data and functionality would have landed rather differently at a non-native company, for example.  Even at Amazon, though, continuous digital transformation is challenging, as \citet{yegge:2011:steveys-google-platforms-rant} relates in his account of the aftermath of the Bezos email.}  In 2018, nearly 10 years after internet giants had begun using machine learning at scale (e.g. Amazon's random forest demand forecasting (2009) and Google's Panda search algorithm (2011)), AI began to gain traction at other firms (\vref{figure:ai-adoption-over-time}).\footnote{\citet[157]{bresnahan:2019:artificial-intelligence-aggregate-growth-prospects} describes the use of machine learning at Amazon, Facebook, Google, and Netflix, particularly for matching (e.g. consumers to products) and for user interfaces and notes ``there is little application of [artificial intelligence technologies] outside the Internet Giants as of spring 2018.''  \citet[1]{bughin-vanzeebroeck:2018:artificial-intelligence-digital-base-critical} notes that ``only a fraction of companies---about 10 percent---have tried to diffuse AI across the enterprise, \ldots An additional quarter of companies have tested AI to a limited extent.''}  Digital natives will surely lead the charge for genAI as well.  For other firms, the pace and success of knock-on innovation will be a key determinant of the scale and timing of productivity effects from genAI.\footnote{The emergence of cloud services may have catalyzed machine learning adoption outside of digital natives, and AI as a service such as Azure OpenAI Service may play a similar role for genAI.}

\subsubsection{Products}\label{section:products}

\paragraph{User interfaces} (UIs) provide a channel through which requests and responses can pass between the user and the model, whether a human user or another system, such as a vehicle or a robot.  In the early days of genAI, users accessed genAI through their own Python programs or through websites such as the OpenAI Playground. A major shift occurred in November 2022, when OpenAI released ChatGPT, a conversational interface that made genAI interactions significantly more accessible to a broader audience. Since then, several new interfaces have emerged. In 2023, OpenAI introduced Custom GPTs, enabling users to create specialized LLMs for specific domains, such as LegalGPT for legal matters.\footnote{See ``Introducing GPTs," November 6, 2023. \url{https://openai.com/index/introducing-gpts/}} In 2024, OpenAI announced integration of their ChatGPT model to Apple's Siri voice assistant and Google launched  NotebookLM, which made it easy to upload documents and transform them into interactive discussions.\footnote{See ``OpenAI and Apple announce partnership to integrate ChatGPT into Apple experiences," June 10, 2024. \url{https://openai.com/index/openai-and-apple-announce-partnership/}; See ``NotebookLM gets a new look, audio interactivity and a premium version," December 13, 2024. \url{https://blog.google/technology/google-labs/notebooklm-new-features-december-2024/}}  In addition, there are ``copilots'' that integrate AI into existing user workstreams, notably GitHub Copilot (computer programming) and Microsoft 365 Copilot (office productivity).  

\paragraph{System interfaces} allow hardware and software  systems to access the core AI system.  For example, Nvidia's Isaac Software Development Kit (SDK) facilitates the integration of AI into robotics.\footnote{Robot capabilities have advanced in tandem with AI progress.  Industrial robots were trained using machine learning beginning in the 1990s \citep{arinez-etal:2020:artificial-intelligence-advanced-manufacturing-current, soori-arezoo-dastres:2023:artificial-intelligence-machine-learning-deep}.  Even more sophisticated robots, which can learn from their environments, integrating sensor data, are available today, including civilian and military autonomous vehicles \citep{knight:2016:japanese-robotics-giant-arms-brains, pierson-gashler:2017:deep-learning-robotics-review-recent}.}  Access to AI through SDK helps the robot with environmental integration problems, such as simultaneously tracking its location and mapping its environment (SLAM).  Development of multimodal models which can take in inputs of different kinds (text, images, sensor readings) and can output instructions to the robot, such as the rotation and torque for a joint, have pushed robot-AI integration forward \citep{reed-etal:2022:generalist-agent, brohan-etal:2023:rt2-vision-language-action-models}

System interfaces also make possible the design of agentic AI systems, which collect information during operation, interpret context, make decisions and act in pursuit of goals autonomously \citep{park-etal:2023:generative-agents-interactive-simulacra-human}.\footnote{For more detail on agents, see \citet{wiesinger-marlow-vuskovic:2024:agents}. For an array of agentic AI use cases, see \citet{renner-chabanl:2024:601-real-world-genai-use}.}  Examples include AutoGen, from Microsoft, and CrewAI.  For example, an airline reservation app using agentic AI would be capable of handling the complex steps of searching for available flights, proactively aligning the user's travel options with their preferences, making the booking, adjusting to disruptions to the reservation process (e.g. sporadic price changes), and including special accommodations along the way (e.g. upgrading seats if available within the user's preferences).

\subsubsection{Production processes} 

\paragraph{Product design} is a use of genAI that will be self-evident to users of tools such as DALL-E to create whimsical images; powerful genAI tools are also capable of designing products of all kinds that meet technical and aesthetic specifications. Perhaps less obviously, the design process itself can be transformed through knock-on innovation.  \citet{saadi-yang:2023:generative-design-reframing-role-designer} interviewed designers and observed:
\begin{quote}
Rather than thinking about how to create several one-off designs, designers may consider how to create a system for design that would allow the design tool to generate a large number of valid outputs. This can involve setting the appropriate specifications, manufacturing methods, and product architecture early in the
process to input into computational tools.
\end{quote}

\paragraph{Production line operation} \citet{serradilla-etal:2022:deep-learning-models-predictive-maintenance-survey} provides an overview of the use of deep learning, including genAI such as generative adverserial networks, to optimize line configuration, throughput, efficiency, and carbon footprint.  Predictive maintenance using synthetic data and scenario simulation is another application for genAI in industry. \citet{sai-sai-chamola:2024:generative-ai-industry-50-analyzing} provides examples of production optimization outside of manufacturing as well.

\subsubsection{Organization} Among the organizational innovations spurred by AI are cross-functional teams with access to data that spans the enterprise, breaking down barriers between business units, optimization of supply chains, and reallocation of employees to de-emphasize repetitive writing tasks \citep{iansiti-lakhani:2020:competing-age-ai-strategy-leadership}.

\subsubsection{Case study evidence} Task-specific user interface programs are already in use in the \textbf{health care} industry. For example, one hospital system used genAI to field a flood of questions during the COVID-19 pandemic \citep{wittbold-etal:2020:hospitals-using-ai-battle-covid19}.  AI is also used to transcribe conversations and dictations to create clinical notes \citep{handa-sorensen:2023:automatically-create-clinical-documentation-generative}.  \citet{hornback-etal:2025:fhir-focus-enabling-biomedical-data} document how Fast Healthcare Interoperability Resources (FHIR), a protocol for the secure exchange of sensitive health information, has evolved to support emerging AI technologies.  These efforts include the integration of BERT, a transformer model, for encoding unstructured text data \citep{Peterson-etal:2020:corpus-driven-standardization-framework-encoding} and the development of a generative model, FHIR-GPT \citep{li-etal:2024:fhir-gpt-ehances-health-interoperability}.   In \textbf{finance}, JP Morgan Chase adopted AI for contract review using software called COiN (Contract Intelligence) and reported saving 360,000 hours of costly lawyers and loan officers \citep{weiss:2017:jpmorganchase-uses-tech-save-hours}.  In the \textbf{information sector}, \citet{peng-kalliamvakou-cihon-demirer:2023:impact-ai-developer-productivity-evidence} document that the use of GitHub Copilot markedly increased the productivity of programmers, particularly less experienced ones.  And, data center networks have evolved to better support genAI and genAI has contributed to network optimization beyond the contributions of machine learning \citep{liu-etal:2024:generative-ai-data-center-networking}.  GenAI has proved useful in the \textbf{energy sector} as well.  \cite{choi-etal:2024:egridgpt-trustworthy-ai-control-room} describe a custom interface to genAI designed to assist control room operators in balancing supply and demand on electrical grids.  

\subsection{Ongoing Core Innovation}

\begin{figure}[ht]\caption{Number of Parameters and Training Dataset Size}\label{figure:number-parameters-training-dataset-size}
	\centering
    \includegraphics[scale=0.37]{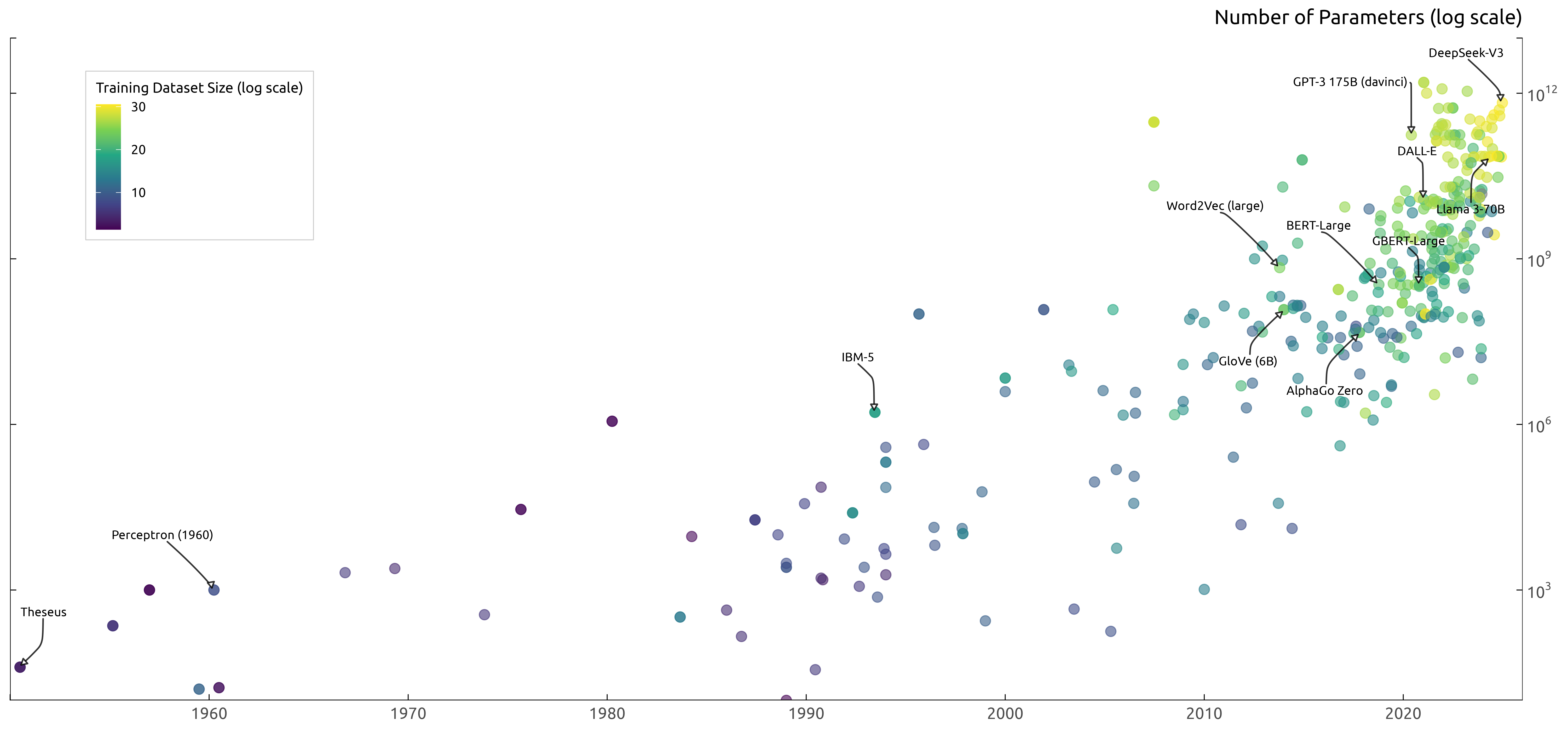}
\begin{minipage}{\linewidth}\setlength{\belowdisplayskip}{0pt} \setlength{\abovedisplayskip}{0pt}
\scriptsize
		{\vspace{2em}Note: Only models with reported parameter size and training dataset size are included in the dataset. For instance, GPT-4 is excluded as its parameter size is not known.\\Source: Epoch (2024) with major processing by Our World in Data \citep{rahman-owen:2024:tracking-compute-intensive-ai-models}.}
\end{minipage}
\end{figure}

Since the introduction of the Transformer, genAI models have steadily pushed out the frontier of capability. At first, advances were largely driven by increasing the scale of the model (via the number of ``parameters" in the model), computational power used (``compute'', in the lingo of AI), and the size of the training dataset (\vref{figure:number-parameters-training-dataset-size}).  AI scientists and engineers often focus on the ``scaling laws'' that describe the benchmark performance effects of increasing each of these inputs \citep{kaplan-etal:2020:scaling-laws-neural-language-models}. More recently, tactics to use compute more efficiently and to refine models for specific applications have been a focus as well.

Importantly, \textit{economic} performance (productivity) only rises when more can be accomplished \textit{while holding input costs fixed}.  In other words, we are looking for \textit{shifts} in scaling law functions, not movement along the curves.  Accordingly, we focus below on (a) how innovations in model architecture (the algorithms used for distilling information from data) raise genAI model capabilities without raising training costs, (b) how hardware innovations lower the cost of computation, and (c) how richer datasets (with more information per token) can be brought to bear on training.

\subsubsection{Model Development}\label{section:model-development}

\begin{figure}[ht]\caption{Open-Source AI Models}\label{figure:open-source-ai-models}
	\centering
	\includegraphics[scale=0.6]{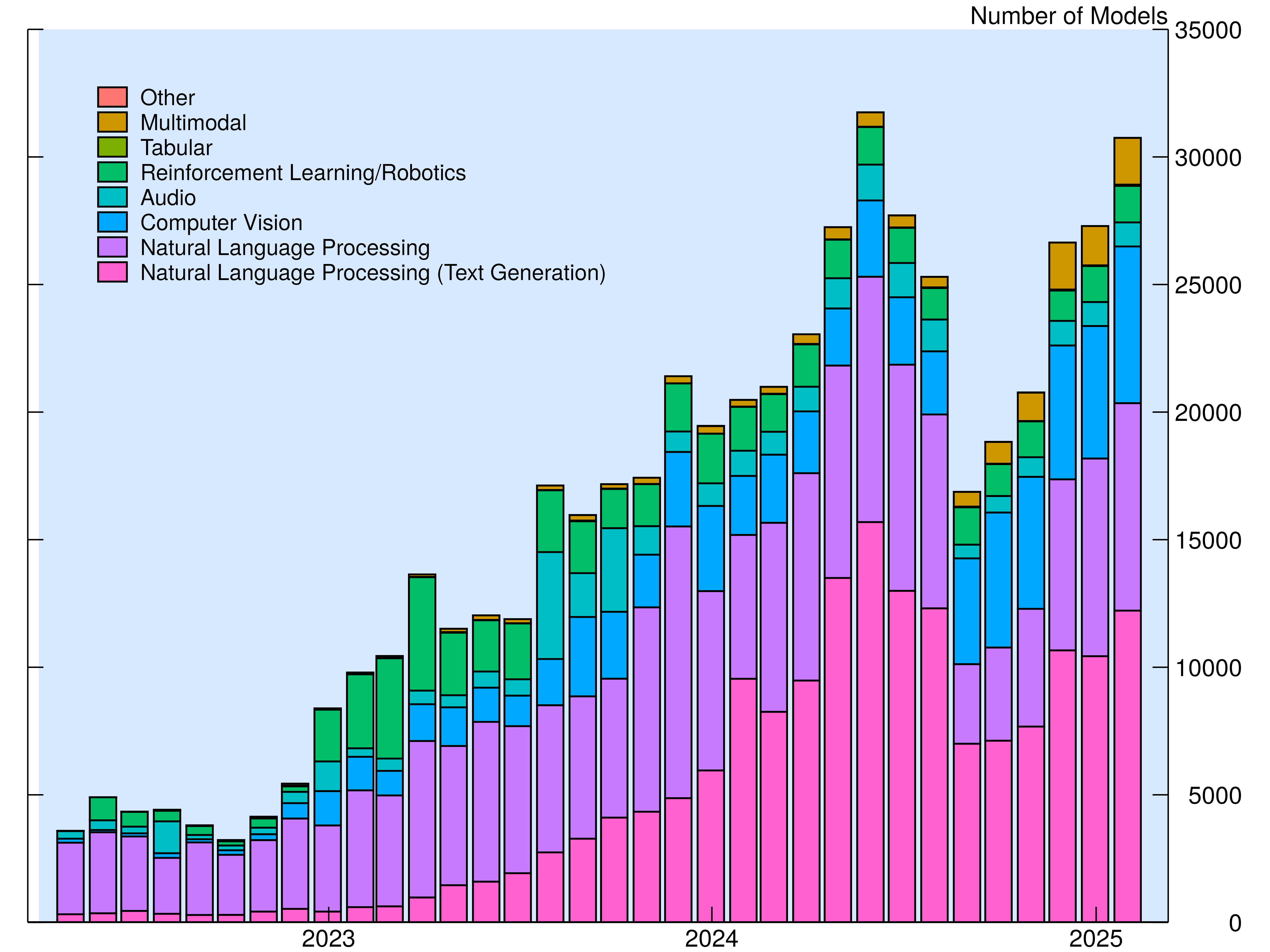}
\begin{minipage}{\linewidth}\setlength{\belowdisplayskip}{0pt} \setlength{\abovedisplayskip}{0pt}
\scriptsize
		{\vspace{-2em}Source: Hugging Face.}
\end{minipage}
\end{figure}

Since the introduction of the Transformer, model development has progressed at a blistering pace, including improvements attributable to ramping up model scale, training dataset size, and the amount of compute employed---the scaling laws discussed above---and the introduction of novel model concepts and techniques to increase the efficiency of model training. An important catalyst in this process has been a rise in open-source models, which has accelerated experimentation across text-generation models, and since the second half of 2024, multi-modal models (\vref{figure:open-source-ai-models}).

The training of genAI models (optimal calibration of its parameters) takes place in two stages: pre-training and fine-tuning.  Pre-training produces a broadly-applicable ``foundation model''; fine-tuning refines the foundation model for a specific application.  The trained model is then used in inference---responding to user requests. Efforts early in the wave of genAI improvement that followed the Transformer focused on pre-training, but the escalating cost of making progress in that stage has led researchers to explore improvements in fine-tuning and inference, as discussed below.\footnote{See Zeff, Maxwell. ``Current AI Scaling Laws Are Showing Diminishing Returns, Forcing AI Labs to Change Course.'' TechCrunch (blog), November 20, 2024. \url{https://techcrunch.com/2024/11/20/ai-scaling-laws-are-showing-diminishing-returns-forcing-ai-labs-to-change-course/}.}  

In addition to training and inference innovations, advances in performance have come from novel model concepts.  Mamba, introduced in 2023, achieved subquadratic-time sequence modelling by avoiding the pairwise comparison among tokens used in the attention mechanism, meaning that as input texts lengthen, the computational burden increases at a slower pace than the Transformer \citep{gu-dao:2023:mamba-linear-time-squence-modeling}.\footnote{In particular, Mamba estimates the parameters of a latent state space structure.}  Small-scale models, with lower computational requirements, have been a focus for some applications as well, such as personal devices and lower resource settings, making LLMs more accessible to the average user.  Microsoft's Phi and models from Mistral AI, in particular, have shown relatively strong performance given their size \citep{jiang-etal:2023:mistral-7b, abdin-etal:2024:phi-3-technical-report-highly-capable}. 

\paragraph{Pre-training} Between 2018 and 2022, a key pre-training tactic in the effort to improve genAI performance was to increase model size.  Size is measured in the number of estimated parameters, most notably the weights that determine how much influence each neuron has on each of the others in a neural network. For example, GPT models initially had 117 million parameters in 2018 (GPT-1), then 1.5 billion in 2019 (GPT-2), and a staggering 175 billion in 2020 (GPT-3).\footnote{Shree, Priya. ``The Journey of Open AI GPT Models.'' Walmart Global Tech Blog (blog), November 10, 2020.\url{https://medium.com/walmartglobaltech/the-journey-of-open-ai-gpt-models-32d95b7b7fb2}} Unfortunately, costs typically rise quadratically when parameters are added: each word (token) in the input sequence has to be compared to all of the others in the attention mechanism, as discussed in the box on the Transformer. 

Remarkably, the cost of training a model of a given size was halved approximately every eight months through 2024 because of improvements in algorithmic efficiency \citep{ho-besiroglu-erdil-owen-rahman-guo-atkinson-thompson-sevilla:2024:algorithmic-progress-language-models}. But, by 2022, the direction of innovation had begun to shift; scaling laws indicated diminishing returns to model size for foundation models, and focus turned to optimizing training efficiency. \cite{hoffmann-etal:2022:training-compute-optimal-large-language}, for example, note that raising the number of parameters faster than the amount of data used leads to ``undertrained'' (imprecisely estimated) parameters that are less useful for inference and recommend scaling the model size no faster than the dataset size.
 
\paragraph{Fine-tuning}

As the returns to model size scaling have diminished, researchers have begun focusing more attention on fine-tuning foundation models for specific tasks. That is, developers have used domain-specific training data to increase the model's expertise beyond the capabilities of foundation models for narrowly defined questions. 

Several techniques have been developed to improve this type of foundation model adaptation.  \textbf{Transfer learning} involves taking a foundation model that has been fine-tuned for a specific task and adapting it (fine-tuning it again) for a related task, a process which typically involves only a modest amount of modification. \textbf{Instruction tuning} provides the model with guidance for specific scenarios that enhance its performance for targeted use cases. For example, the Alpaca model is a fine-tuned version of Meta's Llama model, refined by adding a set of instructions and desired outputs to the training process \citep{taori-etal:2023:stanford-alpaca-instruction-following-llama}.\footnote{The supplemental training set showed Llama the desired response to particular queries, such as ``Instruction: Brainstorm a list of possible New Year's resolutions.  Output:  lose weight, exercise more, eat healthier.''}  \textbf{Reinforcement learning} has been extensively applied in AI for fine-tuning, allowing the model to refine its behavior based on a reward function \citep{mnih-etal:2013:playing-atari-deep-reinforcement-learning}. A particularly influential approach along these lines was \textbf{reinforcement learning from human feedback (RLHF)}, a technique that aligns the model's output with human preferences by learning explicitly from human reactions to the model's responses to their queries \citep{christiano-etal:2017:deep-reinforcement-learning-human-preferences}. For example, the model may provide the user with several responses to a query and ask the user to rank them, then use the ranking to improve performance in future queries.  This technique gained widespread attention in the context of genAI with the release of InstructGPT in 2022 \citep{ouyang-etal:2022:training-language-models-follow-instructions}. 

\paragraph{Inference}
Inference costs (in terms of electricity, time, compute, and carbon emissions) have risen with the popularity of genAI, leading to a focus on techniques to make this step more efficient.\footnote{After setting a flat fee for access to ChatGPT, OpenAI CEO Sam Altman was surprised by the flood of user requests. See ``Sam Altman says he’s losing money on OpenAI’s \$200-per-month subscriptions: `People use it much more than we expected'.''  Economic models have long predicted this outcome when the price of the marginal unit is set to zero.  See, for example, the discussion of overuse of common pasture in \textit{The Wealth of Nations} \citep{smith:1776:inquiry-nature-cause-wealth-nations}.}  \begin{itemize}
    \item One of the most important innovations in this area has been the \textbf{Mixture of Experts (MoE)} approach, an architecture that activates only a subset of model parameters in response to queries \citep{jacobs-etal:1991:adaptive-mixtures-local-experts}. A router determines which ``expert'' (portion of the full model) to activate based on the input. This adjustment allows the model to employ only a subset of the billions of parameters in the original foundation model, drastically reducing computational costs \citep{shazeer-etal:2017:outrageously-large-neural-networks-sparsely}. \
    \item \textbf{Pruning} is another inference refinement; here extraneous parameters are removed outright from the model \citep{cetin-etal:2024:evolved-universal-transformer-memory}.  
    \item Developers also incorporate \textbf{distillation}, a compression-like technique that uses knowledge from large complex models to inform smaller, less costly models \citep{hinton-etal:2015:distilling-knowledge-neural-network}. 
    \item \textbf{Quantization} reduces the level of accuracy in order to reduce costs in terms of computation and memory requirements (e.g. moving from 32-bit to 8-bit floating point precision). For example, in 2023 Microsoft developed BitNet, an LLM with competitive performance that transforms floating point parameters in the model to a terniary digit (0, 1, or -1) \citep{wang-etal:2023:bitnet-scaling-1-bit-transformers}.  
    \item \textbf{Token caching} involves temporarily storing information anticipated to be needed in future inference steps.\footnote{The fact that caching, a technique first employed in computing in the 1960s (on IBM System/360 mainframes), was first introduced to genAI inference in 2023 suggests there may be other basic numerical methods that have yet to be leveraged.  If so, this bodes well for future efficiency improvements.}  For example, the prompts sent by a user to a model for inference may tend to be similar, implying that by caching processed tokens (words, phrases), anticipated computation costs can be reduced \citep{pope-etal:2023:efficiently-scaling-tranformer-inference}.\footnote{Sakana AI recently introduced an approach using neural networks called NAMMs to optimally decide whether to keep or discard tokens, saving up to 75\% of cache memory.  See Dickson, Ben. ``New LLM Optimization Technique Slashes Memory Costs up to 75\%.'' VentureBeat (blog), December 13, 2024. \url{https://venturebeat.com/ai/new-llm-optimization-technique-slashes-memory-costs-up-to-75/}.}    
\end{itemize}
The recently introduced DeepSeek R1 model leverages several of the techniques described above to deliver a substantial performance improvement compared to existing models (See the box, ``Landmark AI Models: DeepSeek R1.'').

In some cases, recent efforts have acted to extended inference time to enhance performance.  Extending inference time raises costs, but on the margin, resources may be better used for responding to queries than for additional training.  OpenAI's recent o1 model exemplifies the benefits of this approach via its superior performance across various domains, particularly reasoning-heavy tasks.\footnote{See Nu\~{n}ez, Michael. ``OpenAI Scientist Noam Brown Stuns TED AI Conference: '20 Seconds of Thinking Worth 100,000x More Data'.'' VentureBeat (blog), October 23, 2024. \url{https://venturebeat.com/ai/ openai-noam-brown-stuns-ted-ai-conference-20-seconds-of-thinking-worth-100000x-more-data/}.}  And, constraining the model to provide a response grounded in logic can lead to better results as well; \textbf{chain-of-thought (CoT) reasoning} guides the LLM to articulate a series of steps in its inference \citep{wei-etal:2022:chain-thought-prompting-elicits-reasoning}.

\begin{tcolorbox}[every float=\centering, breakable,enhanced,title = Landmark AI Models: DeepSeek R1, title after break = DeepSeek R1 (continued), parbox = false]\label{box:deepseek-r1}A salient example of recent model innovation occurred in January, 2025 when DeepSeek unveiled R1 (short for ``reasoning'') \citep{deepseekai:2025:deepseek-r1-incentivizing-reasoning-capability}.  This model set a new bar for cost-effective, high-quality multi-modal models.  DeepSeek stunned the AI research community by reporting the costs for training this 671 billion parameter model to be just under \$6 million.\footnote{See the technical report for DeepSeek V3 \citep[5]{deepseek:2024:deepseek-v3-technical-report}: ``...DeepSeek-V3 costs only 2.788M GPU hours for its full training. Assuming the rental price of the H800 GPU is \$2 per GPU hour, our total training costs amount to only \$5.576 million. Note that the aforementioned costs include only the official training of DeepSeek-V3, excluding the costs associated with prior research and ablation experiments on architectures, algorithms, or data.''} Although observers noted this figure does not account for the substantial reliance on R\&D by other AI developers, the news led to a material change in perceptions of the role of Chinese AI companies.  News of DeepSeek R1 coincided with a one-day market valuation drop of nearly \$600 billion for NVIDIA as DeepSeek had achieved leading edge performance with far less of NVIDIA GPU-provided compute than previously believed necessary.\footnote{See Subin, Samantha. ``Nvidia Sheds Almost \$600 Billion in Market Cap, Biggest One-Day Loss in U.S. History.'' CNBC, January 27, 2025. \url{https://www.cnbc.com/2025/01/27/nvidia-sheds-almost-600-billion-in-market-cap-biggest-drop-ever.html}.} And, the DeepSeek R1 release may have spurred a major competitor to accelerate their product offerings: OpenAI soon updated o3-mini, its extended inference model, lowered its price, and offered a free-trial version.

Deepseek R1 blends several concepts that had been known in the genAI field for some time to improve performance and slash inference costs including mixture of experts, chain-of-thought reasoning, reinforcement learning, distillation, and quantization.  Deepseek R1 endeavored to optimally combine all of these techniques to reduce costs. Some of these techniques were already commonly used in frontier models but others, such as their novel approach to reinforcement learning, called Group Relative Policy Optimization (GRPO), were not.

Relative to other frontier model developers, DeepSeek shifted attention from supervised learning during the fine-tuning stage to reinforcement learning.
\end{tcolorbox}

\paragraph{Agents}\label{agents}  Another direction for progress currently receiving intense attention---distinct from the pre-traning/refinement/inference optimization approach---is the creation of AI agents. \textbf{Agentic AI} systems develop strategies to pursue broad goals and recalibrate in response to their environment, in contrast to \textbf{tool-based AI}, which has a stable structure and calibration and is equipped only to respond to carefully crafted requests.  While agents of different kinds are commonplace in the home and at work, both in physical form, such as self-driving cars, and in virtual form, such as web browsers, agentic AI is distinguished by its autonomy in pursuit of more abstractly specified goals.  One definition comes from \textit{Boston Consulting Group}:\footnote{See Boston Consulting Group. ``AI Agents.'' Accessed August 1, 2025. \url{https://www.bcg.com/capabilities/artificial-intelligence/ai-agents}.}
\begin{quote}
Put simply, AI agents are artificial intelligence that use tools to accomplish goals. AI agents have the ability to remember across tasks and changing states; they can use one or more AI models to complete tasks; and they can decide when to access internal or external systems on a user’s behalf. This enables AI agents to make decisions and take actions autonomously with minimal human oversight.    
\end{quote}

AI agents extend capabilities of AI to wider use in business, particularly to people with little technical knowledge or skill. For example, in the case study of health care, we noted that LLMs are being used to help with the paperwork burden faced by health care professionals, including making appointments, following up on treatment protocols and submitting insurance claims. Specialized AI agents can be programmed to provide this rather specific type of assistance, operating in combination with LLMs.

However, agents require programming and testing and cannot simply apply an off-the-shelf AI program. Correctly specifying the objective function for the AI agent is difficult.  It may inadvertently be guided to maximize its programmed reward while subverting the real objective of the user.\footnote{See ``Faulty Reward Functions in the Wild,'' February 14, 2024. \url{https://openai.com/index/faulty-reward-functions/}.} Moreover, in extreme cases bad actors may hijack agents in business or military uses, leading to disastrous outcomes. 

\subsubsection{Hardware}

GenAI model training and inference has massive computational requirements, making ongoing innovation in electronic hardware (and related hardware, such as cooling systems), essential to continued technical advance.  Progress in this area has been rapid in recent years.  

GenAI processing relies heavily on graphics processing units (GPUs). Like the image processing tasks GPUs were first designed for, training and inference for deep neural networks requires a large number of identical, independent computations which can be run in parallel on a GPU. (In contrast, microprocessing units (MPUs) perform computations sequentially, in the main.) GenAI processing also relies to a lesser extent on field-programmable gate arrays (FPGAs), which are more flexible than GPUs, particularly for inference. And, application-specific integrated circuits (ASICs) are used for specific steps in estimation as well.\footnote{The logical circuits on FPGAs can be reconfigured through programming and tailored to specific algorithms.  The circuitry in ASICs is customized for specific tasks at the time of fabrication and cannot be changed.  Technically, GPUs are ASICs as well.} For example, tensor processing units (TPUs) are customized for matrix multiplication, heavily used in neural networks. They consist of thousands of multipliers and adders connected to each other to form a large physical matrix. Storing the matrix parameters in on-chip registers drastically reduces the need to access off-chip memory, dramatically increasing computational efficiency.

Successive GPUs released by NVIDIA have delivered leaps in AI performance achieved by improvements in the power consumption and computational power of the processing cores---known as Compute Unified Device Architecture (CUDA)---adding and refining TPUs on the GPU chip, and increasing and refining cache (on-chip) memory.   The history of NVIDIA GPU generations illustrates this progression.\footnote{This discussion was substantially improved by a discussion with Claude AI and research on Wikipedia.}  CUDA, first introduced in 2007, explicitly supported non-graphics computing; prior to CUDA, programmers were required to reframe computations in terms of graphics operations.\footnote{With CUDA and associated tools, programmers can use standard programming techniques in \mbox{C/C++}.}  As AI became a prominent use case for NVIDIA GPUs, new architectures were increasingly optimized for deep learning---beginning with Pascal (2016)---and for genAI---beginning with Hopper (2022).  On-chip TPUs were first introduced with the Volta microarchitecture in 2017 and successive GPU generations---Turing (2018), Ampere (2020), Hopper (2022), and Blackwell (2024)---each included improved TPUs.

\begin{table}
    \centering
    \caption{Price of Compute, Selected Nvidia GPUs}
    \label{table:GPU-price-per-tflop}
    \begin{tabular}{|l|l|c|c|c|c|}\hline
        Model & Year & Price  & TFLOPS & Price/TFLOP & Transistors \\ \hline \hline
        GeForce 8800 GT  & 2007 & \$349 & 0.3 & \$1,163 & 0.8B\\ \hline
        GeForce RTX 4060 & 2024 & \$299 & 15.1 & \$20 & 18.9B\\ \hline
        Change (ann. rate) & NA & --1\% & 23\% & --24\% & 19\%\\ \hline \hline
     \multicolumn{6}{l}{Source: TechPowerUp.} \\
    \end{tabular}
\end{table}

\begin{figure}
\centering
\caption{\label{figure:price_indices}Price Indices of GPU Improvements}

\begin{subfigure}[t]{\textwidth}
\centering
\caption{Price per TFLOP}  
\includegraphics[width=0.95\textwidth]{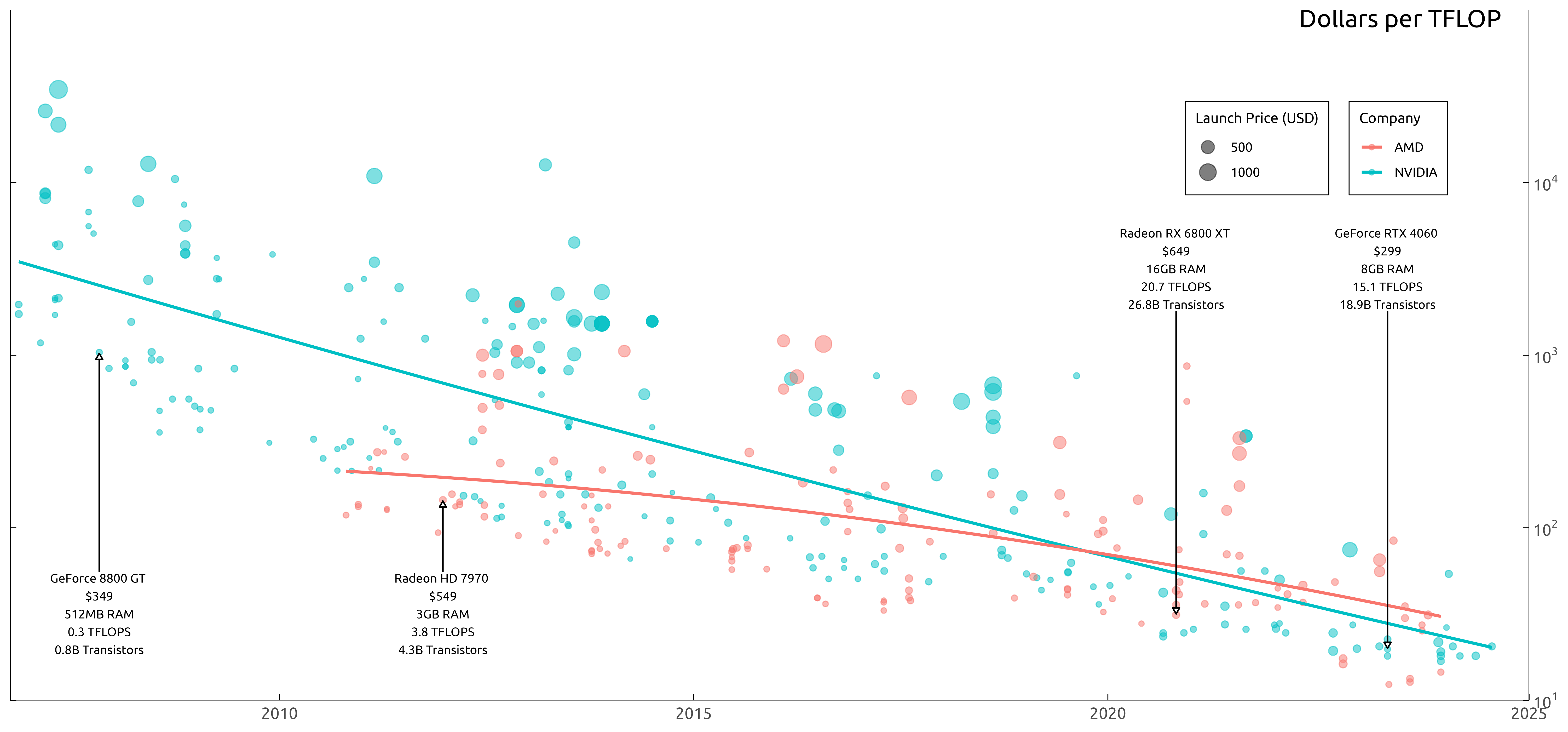}
\caption*{\scriptsize{Note: Price per TFLOPS (trillion floating-point operations per second). The blue line represents the best fit line for NVIDIA GPUs, and the orange line represents the best fit line for AMD GPUs.\\Source: TechPowerUp.}}
\end{subfigure}

\vspace{1cm} 

\begin{subfigure}[t]{\textwidth}
\centering
\caption{Price per vRAM}
\includegraphics[width=0.95\textwidth]{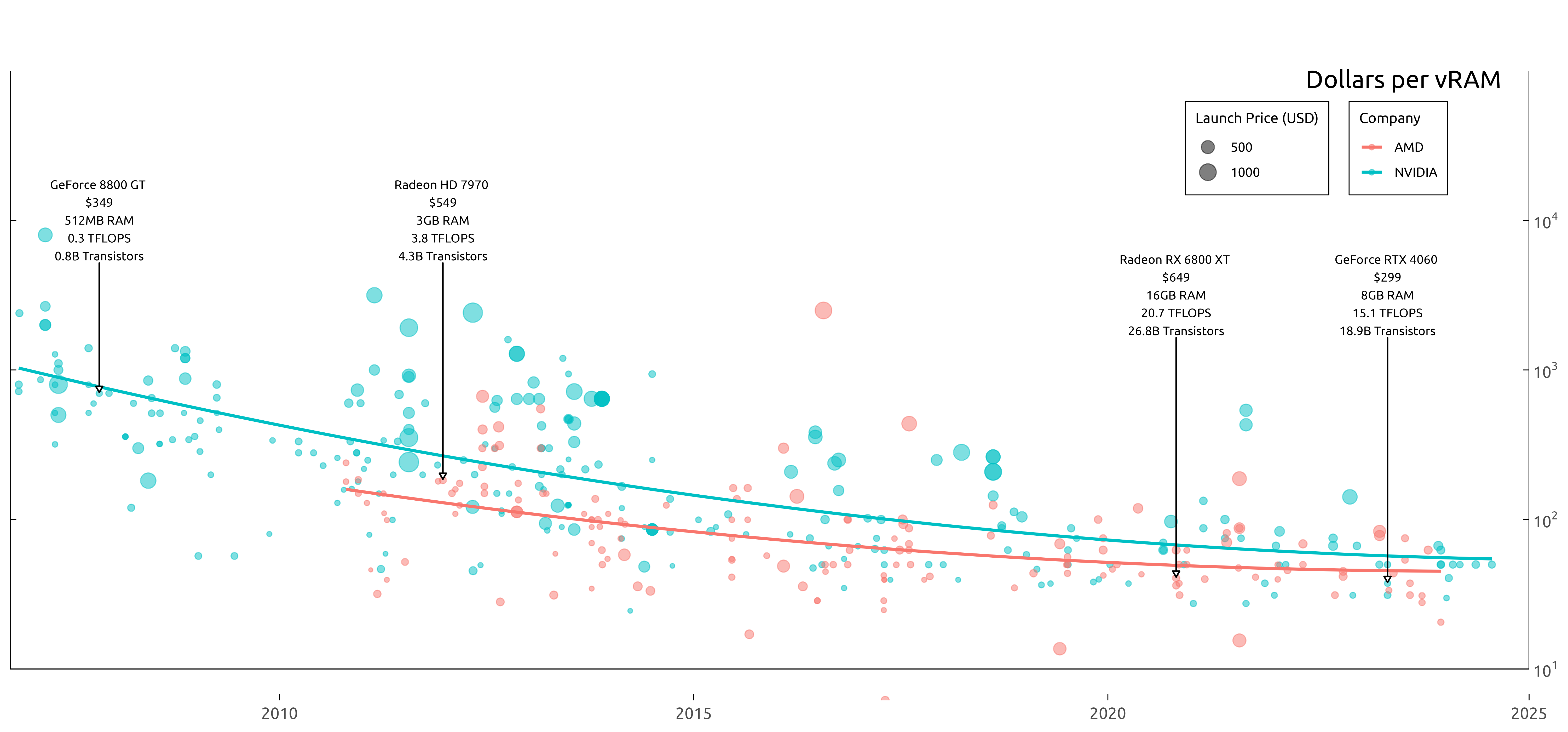}
\caption*{\scriptsize{Note: Price per vRAM in GB (video random access memory). The blue line represents the best fit line for NVIDIA GPUs, and the orange line represents the best fit line for AMD GPUs.\\Source: TechPowerUp.}}
\end{subfigure}

\end{figure}

While the engineering performance of leading edge GPUs has rocketed upwards in recent years, prices have increased dramatically as well.   Fortunately for productivity, holding performance constant, the price of GPUs has moved down:  In 2007, a \$349 GPU provided 0.3 teraflops (TFLOPS) of compute and in 2024, a \$299 GPU delivered 15.1 TFLOPS, implying an average annual rate of price decline of 24\% that persisted for 17 years (\vref{table:GPU-price-per-tflop}).  As shown in \vref{figure:price_indices}, this example is representative of a broader trend:  the cost efficiency of computation has improved dramatically.  Moreover, performance measured in TFLOPS likely understates the advance in AI hardware performance over this period as some design changes operate to reduce the number of TFLOPS needed for a given level of AI training or inference.  For example, GPUs have taken on board additional logic blocks to accelerate matrix operations (Tensor cores) and GPUs have been partitioned to provide more flexible use of TFLOPS \citep{lino:20243:nvidia-gpu-evolution-geforce-ai}.  

\begin{figure}
\caption{Progress on Moore's Law: Two Perspectives}
\centering
\begin{subfigure}[t]{0.48\textwidth}
\caption{\label{figure:moores-law}Transistor Count}
\includegraphics[width=\textwidth]{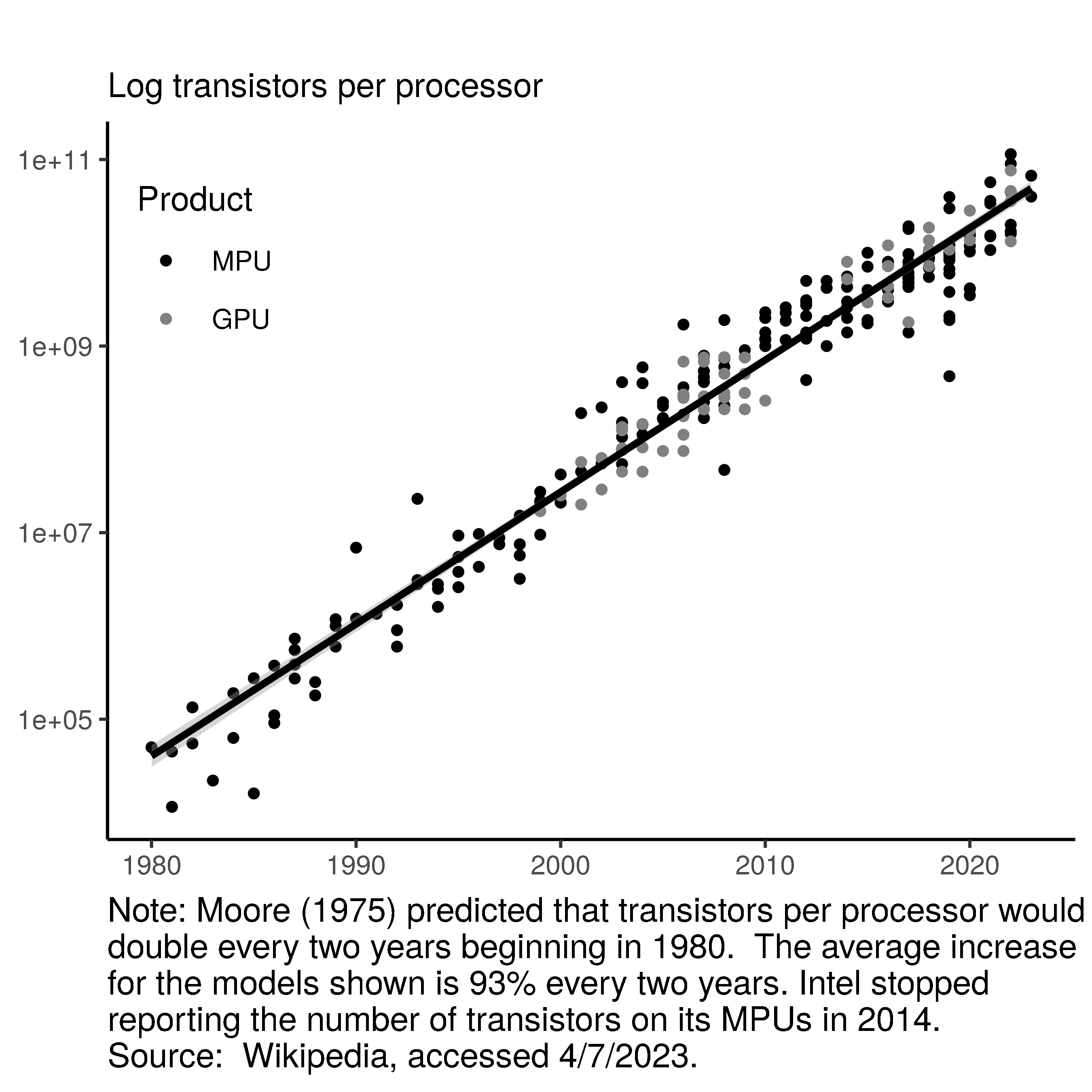}
\end{subfigure}
\begin{subfigure}[t]{0.48\textwidth}
\caption{\label{figure:nodes-no-title}Transistor Size}
\includegraphics[width=\textwidth]{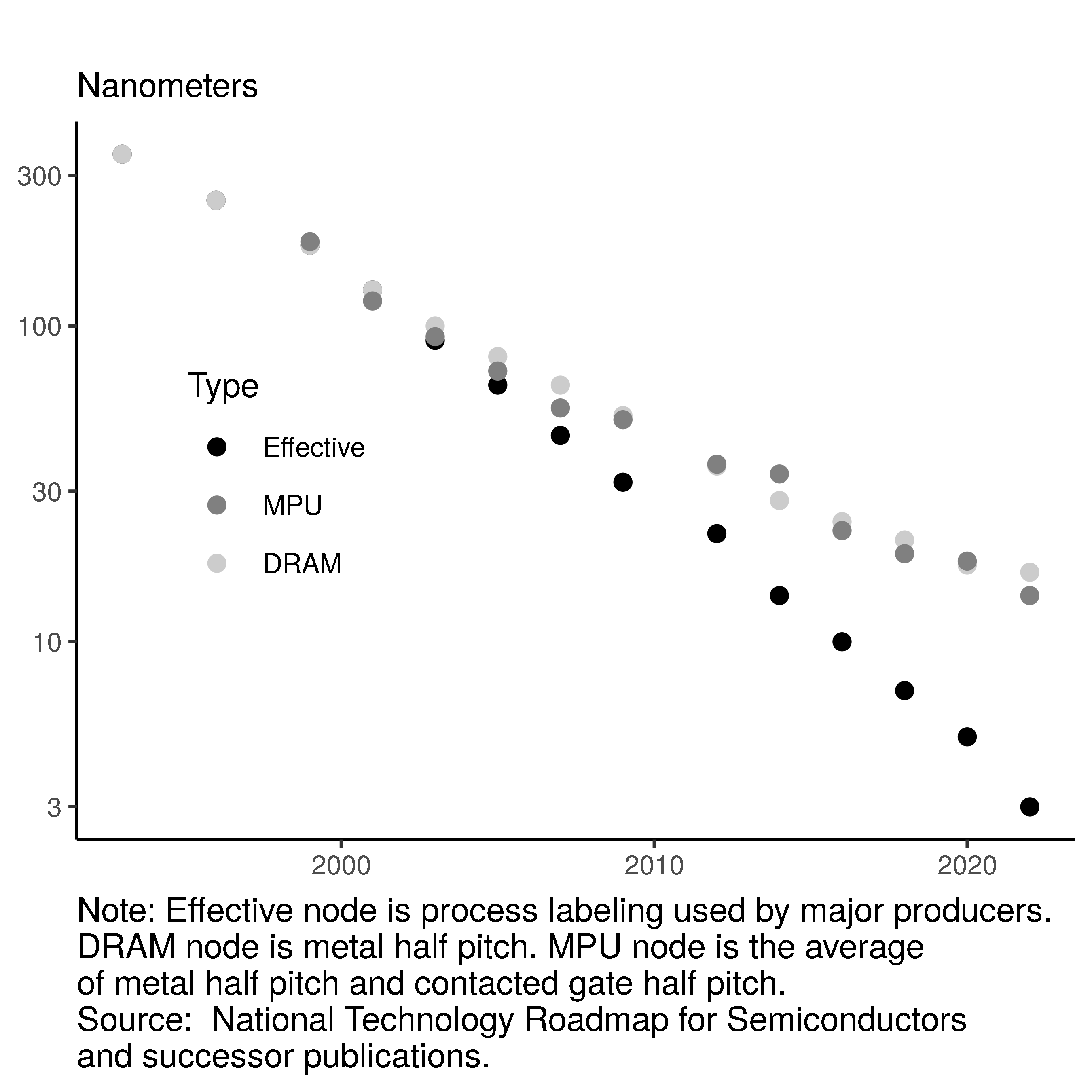}
\end{subfigure}
\end{figure}

Continuation of this trend of declining computation cost is not guaranteed.  What seems like inexorable progress from a distance is in fact the result of a long sequence of difficult engineering feats when seen up close.\footnote{To get a sense of the staggering array of engineering problems involved in moving between chip generations, flip through any edition of the \textit{International Technology Roadmap for Semiconductors} on the worldwide web.}  Historically, a key contributor to falling costs in the semiconductor industry has been steady miniaturization at the leading edge of the chip industry. Through 2003, the linear dimensions of the features (e.g. transistors) on leading edge chips were reduced by 30 percent with each generation (node), yielding a ($0.7*0.7 \approx$) 50 percent reduction in the space they occupied on a chip every two years (\vref{figure:nodes-no-title}). From that year forward, the dimensions of these features were reduced far more slowly, though the industry, using the term ``effective node,'' contended that performance gains with each generation matched the historical trend. The apparent slowdown in cost improvements of TFLOPS and video random access memory (vRAM) in \vref{figure:price_indices} may well reflect that development. Since then, chip innovation has relied less heavily on miniaturization.  For example, increasing ``die size'' (the surface area of the chip) has allowed the number of transistors per chip to continue to climb (\vref{figure:moores-law}).\footnote{Since ``Moore's Law'' is a prediction that the number of transistors per die (chip) will double every two years, it arguably still holds true. Of course, because \citet{moore:1965:cramming-more-components-integrated-circuits,moore:progress-digital-integrated-electronics:1975} does not contain a testable hypothesis, debates about whether Moore's Law holds are largely semantic.  As is evident from \vref{figure:moores-law}, there is no single value for transistors per chip at any point in time.}

\begin{figure}
\caption{Indicators of Power Demand and Supply}
\centering
\label{figure:power}
\begin{subfigure}[t]{0.48\textwidth}
\caption{\label{figure:green500}Data Center Efficiency}
\includegraphics[width=\textwidth]{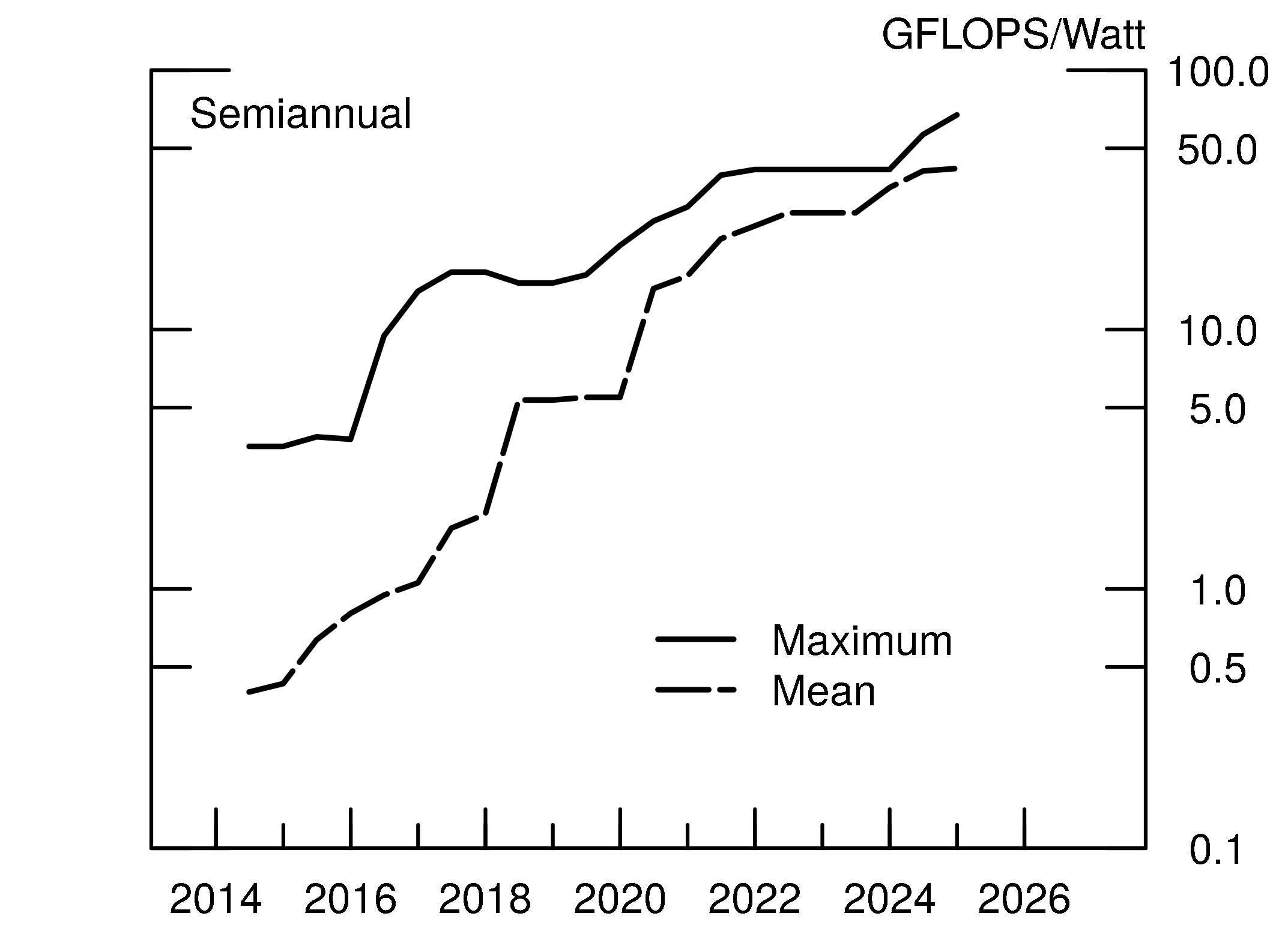}
\end{subfigure}
\begin{subfigure}[t]{0.48\textwidth}
\caption{\label{figure:eia-electricity-forecast}U.S. Electricity Generation}
\includegraphics[width=\textwidth]{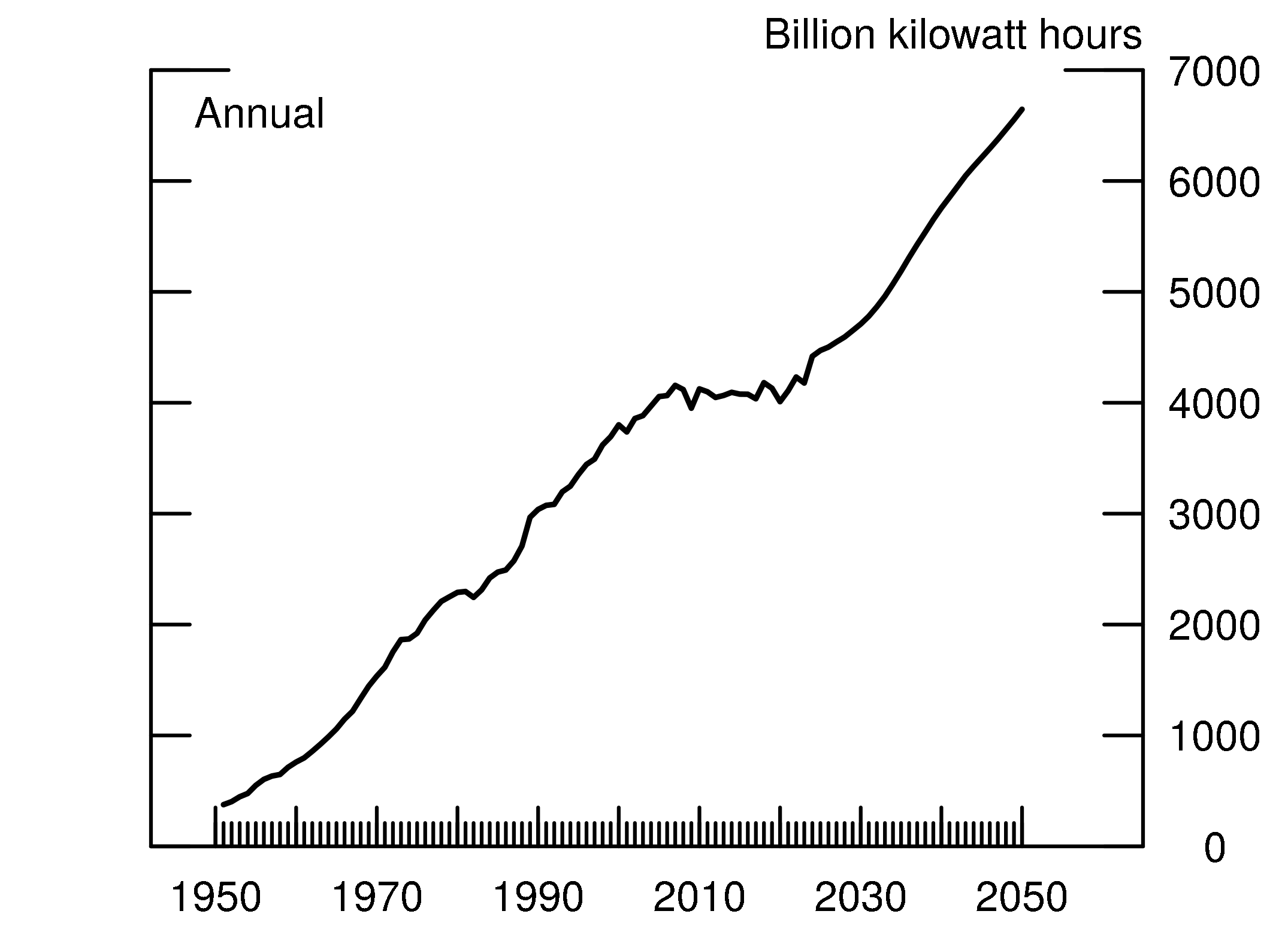}
\end{subfigure}
\begin{minipage}{\linewidth}\setlength{\belowdisplayskip}{0pt} \setlength{\abovedisplayskip}{0pt}
\scriptsize
{\vspace{2em}Note: Data center efficiency indicator is gigaflops per watt of supercomputers labeled ``industry'' or ``vendor''.\\Source: Top500.org for efficiency.  U.S. Energy Information Administration for electricity generation.}
\end{minipage}
\end{figure}

The proximate cause of the miniaturization slowdown was the end of a regularity known as ``Dennard scaling'' whereby power usage was largely unchanged even as more electronic activity was squeezed into the same surface area \citep{dennard-etal:2003:design-ion-implanted-mosfet-small}.  Because heat generation is increasing in power usage, cooling cost began to rise with each node.  Consequently, development efforts in the computing sector shifted to focus on a balance of computing speed and power consumption.  Energy efficiency merits attention as well; concerns have arisen that power demand for genAI use may outpace growth in power-generation capacity, throttling genAI-led productivity advances. Two recent developments temper that concern.  The energy efficiency of the most efficient industrial supercomputers, which include the data centers of major IT service providers, has roughly doubled since 2022, when genAI use began to climb, and U.S. electricity generation is forecast to expand substantially in coming years (\vref{figure:power}).\footnote{Forecasts for the share of power use attributed to data centers in the United States in 2028 range from 6.7 percent to 12 percent, up from 4.4 percent in 2023 \citep{kamiya-coroama:2025:data-centre-energy-use-critical}.}

\subsubsection{Datasets}

GenAI models ``learn'' by adjusting parameters to best represent the content of large amounts of text (and other media), allowing them to estimate the probability that a given word or phrase should appear next in the sequence it generates in response to a prompt. Loosely speaking, the larger the corpus of text available to the model, the better it can estimate these probabilities.  \Vref{figure:number-parameters-training-dataset-size} illustrates the increase over time in the size of the datasets used to train the models.  The change in the coloration of the circles, from purple to blue to green to yellow, indicates the rapid increase in the size of the training datasets.  

Some observers believe that we may be approaching the limit of what we can learn from public text data.\footnote{\citet{villalobos-etal:2022:will-we-run-out-data} predict that the scope of training sets may approach the full extent of public high-quality text data as early as 2026.} A crucial nuance to this aspect of model improvement is that access to more information, not more text in itself, is needed to continue to improve genAI models.\footnote{In terms of information theory, models improve from entropy, the expected amount one will learn from the data generation process producing the text \citep{shannon:1948:mathematical-theory-communication}.}  To illustrate with an extreme example, doubling the size of the training text by exactly duplicating the corpus will yield no improvement to the model. Good quality data enables the model to learn the underlying language structure efficiently, which requires the data to span the full extent of the language with minimal redundancy and noise.  Consequently, there are two looming challenges with respect to training data.  First, more obscure or sensitive topics may have little coverage in public-facing content.  Second, diminishing marginal returns to training will set in as developers move from information-rich content, such as Wikipedia and scientific articles, to more inane text, like social media posts.  

One approach to mitigating the content constraint is transfer learning, where a model pre-trained with public data is improved by further training using proprietary data.\footnote{\citet{cockburn-henderson-stern:2018:impact-artificial-intelligence-innovation-exploratory} note that this raises the issue of market structure as a potential constraint on progress in AI.}  This not only increases the size of the dataset, but may increase the scope as well.  Developers are also wrestling with the question of ``domain generalization,'' where a model is used to generate content on topics outside the scope of the training dataset \citep{zhou-etal:2022:domain-generalization-survey}.

As the information content of the marginal text from the internet falls, other techniques for generating data for training become more attractive.  In one approach, small localized modifications of the training data can be introduced. For example, the performance of an image recognition model may be improved by supplementing the training set of labeled images with their mirror images.\footnote{This approach was taken by the developers of AlexNet, a model which revolutionized the field \citep{krizhevsky-sutskever-hinton:2012:imagenet-classification-deep-convolutional-neural}.}  Such variations in input data constrain the model parameter search process in a useful way.  If an image is a dog, say, the model should recognize that its mirror image is a dog as well, a desirable property known as ``regularity.''  

Another approach to augmentation is the use of ``synthetic data'' created via generative models to emulate the patterns and characteristics of real data \citep{liu-etal:2024:best-practices-lessons-synthetic-data}. For example, an LLM designed to tackle mathematical questions might be trained on a dataset of questions generated by another LLM, using bootstrapping techniques to create similar questions from a human-produced training dataset \citep{yan-etal:2025:s3cmath-spontaneous-selfstep-selfcorrection}. This approach is attractive for medical imagery as well, where creating training data, such as CT scans, is both resource-intensive and constrained by privacy concerns \citep{guo-etal:2025:maisi-medical-ai-synthetic-imaging}.\footnote{The usefulness of synthetic data is a matter of some debate.  Some observers have raised concerns that training with synthetic data (and AI-generated text increasingly present on the internet) will yield low-quality or even nonsensical results, a phenomenon known as ``model collapse'' \citep{alemohammad-etal:2023:self-consuming-generative-models-mad,shumailov-etal:2023:curse-recursion-training-generated-data}. Others have argued that model collapse only occurs when the original training text is replaced by model-generated text \citep{gerstgrasser-etal:2024:model-collapse-inevitable-breaking-curse}.}

Last, datasets can be augmented by harvesting information collected with sensors, particularly in physical environments such as industrial robots and autonomous vehicles \citep{feng-etal:2019:deep-active-learning-efficient-training}.  This approach offers the prospect of a broader domain of use for genAI models and further diffusion of the technology. 

\subsection{The Case that GenAI is a GPT}

To summarize, although it is early to tell how widespread the use of genAI will be, the case that generative AI is a general-purpose technology is compelling, supported by the impressive record of knock-on innovation and ongoing core innovation.

Of the three GPT criteria, widespread adoption is the most difficult to argue that genAI has met. Although some field studies have provided encouraging results that genAI may raise productivity, outside of large corporations, few firms have adopted the technology.  The share of jobs requiring AI skills is low and has moved up only modestly, suggesting that firms are taking a cautious approach.  The ultimate test of whether genAI is a GPT will be the profitability of genAI use at scale in a business environment and such stories are hard to come by at present. That said, use among individuals is high, perhaps unbeknownst to their employers, and with genAI increasingly folded into office productivity software (such as Microsoft 365), its use may become so unremarkable that firms and workers may not be aware it is in use.

The case that genAI meets the knock-on innovation criterion is somewhat stronger.  Key areas of product innovation include user interface software and interface with robotics, where the genAI model enables far more sophisticated applications.  Production process innovation includes digital twins to improve production line efficiency.   Organizational innovations include restructuring of the product design business function.  What share of organizations can justify the digital transformation needed for genAI, such as centralized data governance, particularly for non-digital-native companies, remains to be seen.

Core technology innovation is the criteria for which the case is the clearest.  GenAI performance has moved up at a blistering pace since the introduction of the Transformer thanks to increasingly large datasets and application of more computing power.  More importantly, performance has risen while holding inputs constant (data, compute, and model size), driven by algorithmic improvements. If this trend continues, the direct cost of using genAI will fall, spurring greater adoption.

\begin{tcolorbox}[every float=\centering, breakable,enhanced,title = Artificial General Intelligence, title after break = AGI (continued), parbox = false]\label{box:artificial-general-intelligence}
In 1960, Herbert A. Simon, winner of the 1975 \textit{Turing Award} and 1978 \textit{Nobel Memorial Prize in Economic Sciences} wrote, ``within the very near future---much less than twenty-five years [before 1985]---we shall have the \textit{technical} capability of substituting machines for any and all human functions in organizations'' \citep[22]{simon:1960:corporation-managed-machines}.  This is the concept now known as ``artificial general intelligence'' (AGI).\footnote{Remarkably, \citet{capek:1920:rossums-universal-robots} had already envisioned a world with robots that performed all human tasks, including ones involving physical manipulation of the environment.  At that time, although punchcard tabulators programmable with plugboards existed, most ``computers'' were human \citep{grier:2007:computers-human}.}

Demonstrating the feasibility of AGI is a long-run objective of many AI researchers but is not a particularly interesting exercise for economists.  Technical feasibility is a necessary condition, but far from a sufficient one, for a technology to raise productivity.  It is technically feasible to turn lead into gold, for example, but only at prohibitive cost \citep{matson:2014:fact-fiction-lead-turned-gold}. And, conjecture about future technology, however well informed, is not sufficient for a useful productivity forecast, which must answer the question of when, not just if, the technology will appear and, \textit{a fortiori}, when it will be practical to use.  Moreover, achieving human-level performance on all tasks would likely entail devoting resources to improving performance on tasks AI is ill-suited for at the expense of tasks where improvement would be more readily achieved. \citet{dellacqua-etal:2023:navigating-jagged-technological-frontier-field} explore this issue and provide extensive evidence this is a first-order impediment to AGI.

Some present-day IT leaders believe AGI is imminent: In 2024, Elon Musk predicted the arrival of AGI within two years, Sam Altman predicted its arrival by 2025, and Dario Amodei expected AGI by 2026.  Others are skeptical: Yann LeCun speculated it may never be achieved.\footnote{See \textit{Reuters.} ``Tesla’s Musk Predicts AI Will Be Smarter than the Smartest Human next Year.'' April 8, 2024, sec. Technology. \url{https://www.reuters.com/technology/teslas-musk-predicts-ai-will-be-smarter-than-smartest-human-next-year-2024-04-08/}; Varanasi, Lakshmi. ``Here’s How Far We Are from AGI, According to the People Developing It.'' Business Insider, November 9, 2024. \url{https://www.businessinsider.com/agi-predictions-sam-altman-dario-amodei-geoffrey-hinton-demis-hassabis-2024-11}; PYMNTS.com. ``Meta AI Head: ChatGPT Will Never Reach Human Intelligence,'' May 22, 2024. \url{https://www.pymnts.com/artificial-intelligence-2/2024/meta-ai-head-chatgpt-will-never-reach-human-intelligence/}.}  Historically, technology forecasts have been highly unreliable.  \citet{berkeley:1949:giant-brains-machines-think}, for example, foresaw a machine that would read handwritten text; noteworthy practical use of handwriting recognition systems U.S. Post Office came fifty years later.

Fortunately, AGI is not a precondition for genAI to be a GPT, nor do we need a forecast for the timing of the arrival of AGI to forecast the effects of genAI on productivity.  
\end{tcolorbox}

\section{Is GenAI an Invention of a Method of Invention?}

In classical ``light bulb'' growth models (commonly known as ``Solow-Swann'' models), the source of total factor productivity (TFP)---the part of productivity growth not attributable to capital accumulation---is unspecified \citep{solow:1994:perspectives-growth-theory}.  About the importance of TFP that emerged when this model was brought to the data, \citet[11]{abramovitz:1956:resource-output-trends-united-states} famously observed, ``since we know little about the causes of productivity increase, the indicated importance of this element [TFP] may be taken as some sort of measure of our ignorance about the causes of economic growth in the United States and some sort of indication of where we need to concentrate our attention.'' ``Endogenous growth'' models have since appeared, including some that add a research sector to produce new technologies in response to incentives \citep{romer:1994:origins-endogenous-growth,aghion-howitt:1992:model-creative-destruction,akcigit:2023:creative-destruction-economic-growth}.\footnote{Of course, in actuality TFP is \textit{not} simply the output of a research sector.  TFP results when firms choose, in response to research results, to make complementary investment in intangibles \citep{brynjolfsson-rock-syverson:productivity-jcurve-intangibles-complement-general:2021} in the context of government policy \citep{baily-etal:2025:productivity-policy-united-states} and is importantly affected by business dynamism \citep{decker-haltiwanger-jarmin-miranda:2017:declining-dynamism-allocative-efficiency-productivity}, labor market efficiency \citep{davis-haltiwanger:2014:labor-market-fluidity-economic-performance}, and market structure \citep{goettler-gordon:2011:does-amd-spur-intel-innovate}.}  Like other sectors, efficiency in the research sector can be increased by the use of appropriate capital, such as an invention of a method of invention (IMI).\footnote{The term originates from \citet{whitehead:1925:science-modern-world-lowell-lectures}, according to \citet{mowery-rosenberg:1999:paths-innovation-technological-change-20th}.}  \citet{griliches:1957:hybrid-corn-exploration-economics-technological} noted that the hybridization process developed for creating new varieties of corn played this role. Other examples of IMIs are shown in \vref{table:examples_of_imis}, grouped into observational, analytical, communication, and organizational tools.  We consider below whether genAI falls into these categories and how it can contribute to research productivity beyond what is contributed by machine learning.  We then review a number of broad indicators of the role of AI in research, including patent filings, the share of AI use accounted for by users with research jobs and tasks, and new evidence on the prevalence of AI references in company conference calls.

\begin{table}[ht]
    \centering
    \caption{Examples of Inventions of Methods of Invention}\label{table:examples_of_imis}
    \begin{tabular}{|l|l|}
    \hline
    \multicolumn{2}{|c|}{Observational tools} \\ \hline
     Telescope    & 1608 CE \\
     Compound microscope    & 1620 CE\\
     Pendulum clock    & 1656 CE\\
     DNA sequencer    & 1973 CE \\ \hline
    \multicolumn{2}{|c|}{Analytical tools} \\    \hline
     Mainframe (IBM S/360)  & 1964 CE\\
     Personal computer (IBM PC)   & 1981 CE \\
     Machine learning    & 1998 CE\\ \hline
     \multicolumn{2}{|c|}{Communication tools} \\ \hline
     Printing press (Gutenberg) & 1439 CE \\
     Internet protocol (TCP/IP) & 1975 CE \\ \hline
     \multicolumn{2}{|c|}{Organizational innovations} \\ \hline
     Scientific societies (Accademia dei Lincei)   & 1603 CE\\
     Corporate labs (GE)   & 1900 CE \\
     Government labs (U.S. NRL) & 1923 CE\\
     Big science (Oak Ridge) & 1961 CE\\ \hline
    \multicolumn{2}{|l|}{Source: Authors' judgment.} \\ \hline
    \end{tabular}  
\end{table}

Prior to the appearance of genAI, AI had already diffused across a wide range of scientific disciplines \citep{carobene-etal:2024:rising-adoption-artificial-intelligence-scientific}.   And, it had already been shown to improve the efficiency of research. \citet[23]{cockburn-henderson-stern:2018:impact-artificial-intelligence-innovation-exploratory} note that pre-generative AI assists with the ``labor-intensive search with high marginal cost of search'' involved in many types of R\&D.  Put differently, AI improves prediction, a point emphasized by \citet{agrawal-gans-goldfarb:2018:prediction-machines-simple-economics-artificial}, including predicting how materials might behave.  Examples of phenomenal success are well known.   Scientists have made major advances toward practical nuclear fusion using reinforcement learning techniques to adjust the magnetic system that contains the plasma in a fusion reactor \citep{degrave-etal:2022:magnetic-control-tokamak-plasmas-deep,seo-etal:2024:avoiding-fusion-plasma-tearing-instability}.  \citet{richardson-etal:2020:baricitinib-potential-treatment-2019-ncov} used the ``knowledge graph''---which encodes relationships among scientific publications using machine learning---created by BenevolentAI to produce a novel treatment for COVID-19.  Machine learning has also been used extensively for predicting the properties of novel metal alloys, economizing on physical experimentation and computer simulations \citep{hart-etal:2021:machine-learning-alloys}.  Our focus is on the question of whether genAI enables additional efficiencies in R\&D beyond these and other improvements provided by machine learning.  In particular, we ask whether genAI enhances measurement, analysis, communication, and organization of invention. 

\paragraph{GenAI as an observational tool}
Observational tools, such as microscopes, telescopes, and cameras produce imperfect images due to defects in their components and variation in the environment.  GenAI provides a tool to impute imperfect portions of images as well as missing observations in datasets of all kinds in a fashion more consistent with the apparent properties of the underlying phenomena.  For example, generative techniques for image enhancement, which rely on an implicit model of the manifold of the data generating process---closer to the actual physics, say, of a remote galaxy seen through an imperfect lens---perform better than techniques, such as splines, relying solely on smoothness assumptions (i.e. that nature does not make leaps) \citep{liu-etal:2018:image-inpainting-irregular-holes-partial,lugmayr:2022:repaint-inpainting-denoising-diffusion-probabilistic}.

\paragraph{GenAI as an Analytical Tool}
Like the compound microscope for physical phenomena,  \citet{christian:2020:alignment-problem} notes that LLMs serve as a kind of microscope to look at social phenomena.  \citet[183]{caliskan-bryson-narayanan:2017:semantics-language-corpora-humanlike-biases}, for example, find that ``text corpora contain recoverable and accurate imprints of our historic biases.''  This new visibility may promote and support analysis of social science questions not previously tractable.  There has been an explosion of sentiment analysis and other forms of NLP in recent years fueled by this capability of genAI.\footnote{Sentiment analysis is possible with earlier forms of AI but the capabilities of genAI models are vastly greater \citep{gentzkow-kelly-taddy:2019:text-data,dell:2025:deep-learning-economists}.}  While the identification of underlying sentiment (encoding) is strictly speaking a function of the LLM, conveying the discovered sentiment to the user is necessarily a generative process.   \citet{korinek:2023:generative-ai-economic-research-use} documents a variety of potential roles for genAI in the economic research process; that genAI may play a similar role in many other fields is a reasonable conjecture.\footnote{Early versions of this paper cited \citet{toner-rodgers:2024:artificial-intelligence-scientific-discovery-product}, which purported to show that genAI substantially accelerates the discovery process in materials science.  The veracity of that work has since been questioned by the author's institution and prominent researchers in the field.  Credible empirical evidence on the effect of genAI on scientific research efficiency would be a timely contribution to resolving the uncertainty around the effects of genAI on productivity.}

\paragraph{GenAI-Supported Organizational Innovation}
Institutional organization plays a central role in the effectiveness of R\&D \citep{mowery-rosenberg:1999:paths-innovation-technological-change-20th}, as do informal associations into professional networks \citep{wang-barabasi:2021:science-science} and geographic clusters \citep{porter-stern:2001:innovation-location-matters}.  Consequently, the method of invention for any given research program properly includes the institutions involved.  Emerging applications of AI ``digital twins'' offer the prospect of R\&D with a reduced institutional footprint in many areas of study.  Among these are drug discovery \citep{bordukova-etal:2024:generative-artificial-intelligence-empowers-digital}, industrial research \citep{tao-zhang-zhang:2024:advancements-challenges-digital-twins-industry}, and materials science \citep{kalidindi-etal:2022:digital-twins-materials}.  For example, generative adversarial networks (GANs) may provide an alternative to animal testing for toxicology \citep{chen-etal:2022:toxgan-artificial-intelligence-approach-alternative}.

\paragraph{GenAI as a Communication Tool}
Although empirical and analytical stages of research projects focus on measurement and calculation, many aspects of the research process involve manipulating language.  GenAI may be employed in the writing tasks involved in the conceptual, planning, and dissemination stages of research projects, such as drafting literature reviews, grant applications, and seminar slides (\vref{table:stages-research-project}).  Whether, on net, genAI improves the efficiency of such tasks once the effort needed for review and editing of the documents drafted by genAI is accounted for is an open question.  If so, genAI may play a similar role to the printing press and word processing as a catalyst to the invention process.

\begin{table}
    \centering
    \caption{The Stages of a Research Project}\label{table:stages-research-project}
    \begin{tabular}{|l|p{9cm}|}\hline
    Conceptual & reviewing the literature, formulating the broad problem,  identifying specific goal \\\hline
    Planning & determining research design and procedures, identifying resource needs, procuring funding \\\hline
    Empirical & collecting data, preparing data for analysis \\\hline
    Analytical & identifying data features, testing hypotheses, interpreting results\\\hline
    Dissemination  & communicating to audience (colleagues, industry, policymakers, public, students) in written, visual, and oral form \\ \hline
    \end{tabular}
\end{table}

\paragraph{Research agents}
AI agents (discussed in \vref{section:products}) have emerged that endeavor to automate the core of research entirely, generating research questions, designing and conducting experiments, and reporting results.  Thus, research agents may play the role of an observational, analytical, and communication IMI all at once.  Examples include Google's AI co-scientist and Sakana's The AI Scientist \citep{gottweis-etal:2025:towards-ai-coscientist,lu-etal:2024:ai-scientist-fully-automated-open-ended}.\footnote{\citet{stoughton:2023:meet-coscientist-ai-lab-partner} documents co-scientist innovation at the National Science Foundation.}  

Opinions of the significance of research agents vary widely.  Importantly, the design, conduct, and communication of experiments is only a portion of the activities of a scientist (\vref{table:other-scientific-activity}).  Even so, \citet{lu-etal:2024:ai-scientist-fully-automated-open-ended} report that The AI Scientist can generate publishable research for as little as \$15 per journal article, a striking finding. On the other hand, \citet[1]{beel-kan-baumgart:2025:evaluation-sakana-ai-scientist-autonomous} evaluate Sakana's agent and conclude, 
\begin{quote}
We evaluated the AI Scientist and found several critical shortcomings. The system's literature review process is inadequate, relying on simplistic keyword searches rather than profound synthesis, which leads to poor novelty assessments. In our experiments, several generated research ideas were incorrectly classified as novel, including well-established concepts such as micro-batching for stochastic gradient descent (SGD). The AI Scientist also lacks robustness in experiment execution—five out of twelve proposed experiments (42\%) failed due to coding errors, and those that did run often produced logically flawed or misleading results. In one case, an experiment designed to optimize energy efficiency reported improvements in accuracy while consuming more computational resources, contradicting its stated goal. Furthermore, the system modifies experimental code minimally, with each iteration adding only 8\% more characters on average, suggesting limited adaptability. The generated manuscripts were poorly substantiated, with a median of just five citations per paper—most of which were outdated (only five out of 34 citations were from 2020 or later). Structural errors were frequent, including missing figures, repeated sections, and placeholder text such as “Conclusions Here”. Hallucinated numerical results were contained in several manuscripts, undermining the reliability of its outputs.
\end{quote}

\begin{table}
    \centering
    \caption{Scientific Activity beyond Research Projects}\label{table:other-scientific-activity}
    \begin{tabular}{|l|p{9cm}|}\hline
    Conceptual &  designing a research program (a connected set of research projects); packaging program to influence appropriate audiences (e.g. writing textbooks)\\\hline
    Leadership  & playing executive and advisory roles in local and profession-wide academic and government specialist communities \\\hline
    Mentoring & supervising research, advising and instructing students\\\hline
    Support & fostering buy-in to research program from institutional leadership \\\hline
    Networking & recruiting collaborators and maintaining relationships \\\hline
    Commercialization & translating research results into practical applications\\\hline
    \end{tabular}
    \label{tab:scientific-activity-beyond-research-projects}
\end{table}

Moreover, genAI research agents may have a subtle but important limitation: uncovering the fundamental features of phenomena.   \citet{li-hopkins-bau-viegas-pfister-wattenberg:2022:emergent-world-representations-exploring-sequence} argue that genAI does have that capability.  They trained a generative model to play the board game Othello without providing the rules of the game then demonstrated that the model can play appropriately in a setting not found in the training data, concluding that genAI has created an ``emergent world model.''  Other research has challenged this conclusion.  \citet{jylin04-etal:2024:othellogpt-learned-bag-heuristics} argue that the model is employing a ``bag of heuristics,'' rather than a set of game rules.\footnote{A useful entry point to this ongoing debate is ``LLMs and World Models,'' by Melanie Mitchell, February 13, 2025, found at the \textit{AI: A Guide for Thinking Humans} Substack blog.}  This question is a crucial one in determining the capabilities of genAI to contribute to science.  Without a model of the underlying structure of the physical or social phenomenon under study, one cannot articulate its fundamental laws.  This limitation may arise naturally from the training process; humans learn the fundamentals of science from textbooks, but these laws may not be the rhetorical foundation for the verbal exchanges on the topic found in the training corpus.

\subsection{Indicators of GenAI Research and of GenAI Use in Research}

We discuss below a set of indicators of genAI research (patents) and of genAI use in research (conference call transcripts and genAI queries).\footnote{For evidence of the potential for genAI use in research based on job descriptions, see \citet[1308]{eloundou-manning-mishkin-rock:2024:gpts-gpts-labor-market-impact} who note that ``scientists and researchers'' and ``technologists'' are the job groups most highly exposed to LLMs, and that ``this suggests that when LLMs improve, they have potential to cause downstream improvements in R\&D productivity for workers in sectors deploying them.''}

\begin{figure}[ht]\caption{AI Mentions in Scientific Patents}\label{fig:patents}
	\centering
	\includegraphics[width=0.8\linewidth]{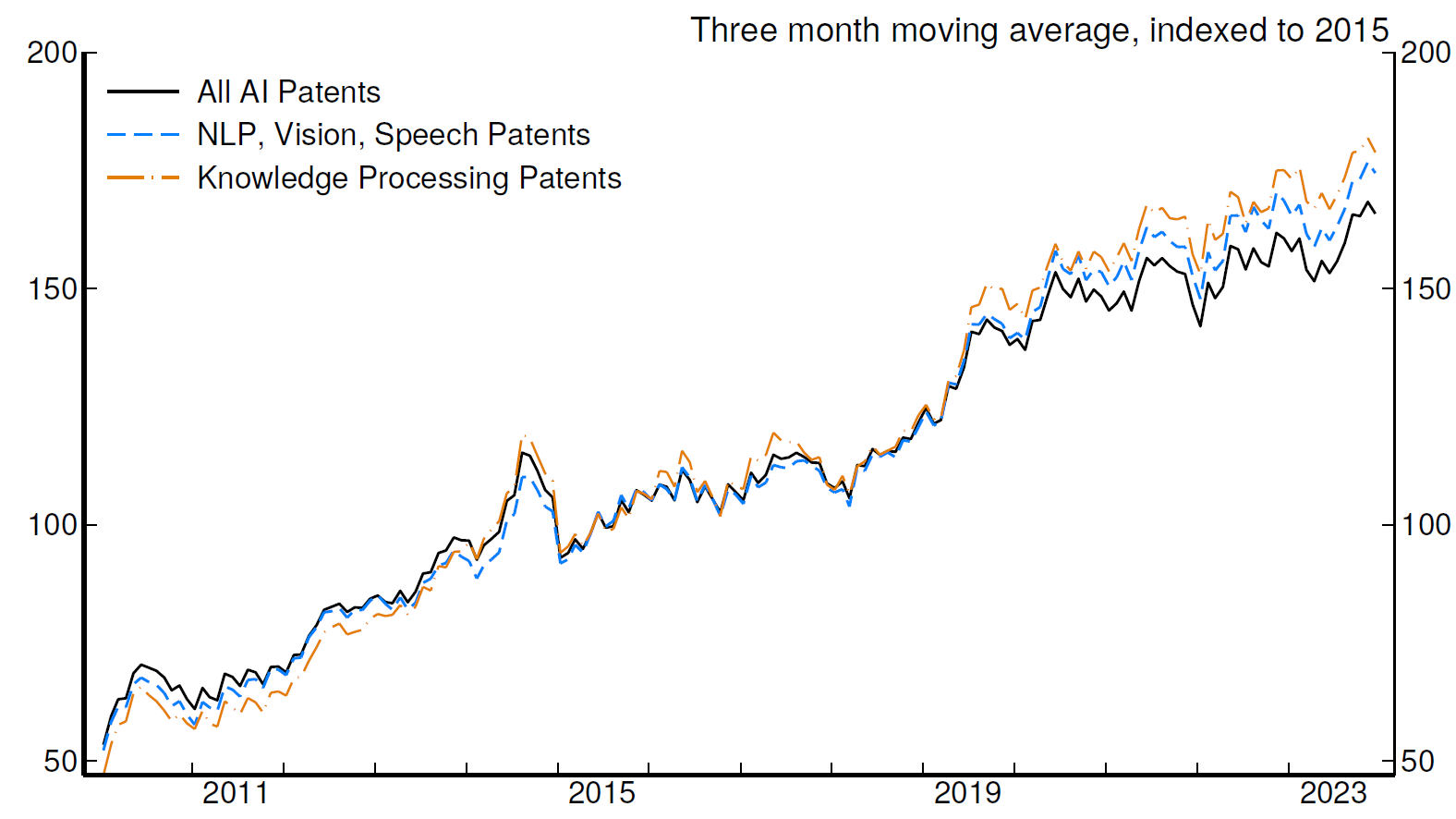}
\caption*{Source: Artificial Intelligence Patent Dataset (2023), U.S. Patent Office.}
\end{figure}

\paragraph{Patents}
AI-related patents issued by the United States Patent and Trademark Office (USPTO) increased markedly following the advent of genAI, suggesting a related surge in genAI research (\vref{fig:patents}).\footnote{\citet{pairolero-etal:2025:artificial-intelligence-patent-dataset-2023} use BERT-based embeddings to refine a previous iteration of their patent classification methodology that used Word2Vec. We use their most conservative threshold of 93\% probability to identify AI-related patents. We refer the reader to the USPTO website hosting their data for more information: https://www.uspto.gov/ip-policy/economic-research/research-datasets/artificial-intelligence-patent-dataset.} The USPTO index of AI-related patents began climbing in 2018, shortly after the publication of the seminal paper by \citet{vaswani-shazeer-parmar:2017:attention-all-you-need} which introduced the Transformer architecture, quickly reaching a level 50 percent higher, which it has sustained since 2019.  We also observe that increases in patent activity for AI modalities particularly related to genAI---natural language processing (NLP), vision, speech, and knowledge processing---have risen even further. This suggests that the recent surge in patenting activity is not merely a reflection of advancements in machine learning.

\paragraph{GenAI Prompts}
\citet{handa-etal:2025:which-economic-tasks-performed-ai} provide a rich set of information on \textit{actual} genAI use in their Anthropic Economic Index (AEI), a useful complement to the detailed work on the \textit{potential} impact of genAI based on analysis of job descriptions from \citet{eloundou-manning-mishkin-rock:2024:gpts-gpts-labor-market-impact}.  The AEI assigns millions of conversations from Claude (Anthropic's premier genAI system) to roughly 3,500 of the tasks defined by the U.S. Department of Labor's O*NET Dataset.\footnote{Patterns of Anthropic use may not be representative of the patterns for all genAI programs, of course.  See section 7.1 of \citet{tamkin-etal:2024:clio-privacy-preserving-insights-real} for a discussion of related work.}  An equal fraction of each task's percentage share of all prompts is then apportioned to each occupation which includes that task in O*NET.

\Vref{table:anthropic1} shows the estimated share of prompts accounted for by occupational groups, their employment share, and the ratio of the two. (If prompts were equally distributed across all workers, these ratios would each be equal to 1.)  ``Computer \& mathematical occupations'', which includes the computer programmers for whom genAI use is especially intense, have the highest ratio of prevalence of genAI use to occupational prevalence, 10.9.  Use intensity is nearly as high among scientists, who account for 7.1 times as many prompts as would be found if prompts were equally distributed across workers.\footnote{Naturally, they may be using genAI for computer programming tasks as well.} Other occupational groups with high relative prevalence of genAI use include ``arts, design, sports, entertainment \& media''; ``architecture \& engineering''; and ``educational instruction \& library''.  The remaining 87.6\% of employment is accounted for by occupations which AEI found had a share of Claude prompts roughly equal to or lower than their share of employment, highlighting the very concentrated nature of genAI adoption in the economy at present.

\begin{table}
    \centering
    \caption{Occupations with High GenAI Usage}
    \label{table:anthropic1}
    \begin{tabular}{|p{6.5cm}|r|r|r|}\hline
        Job Type & Prompt Share & Empl. Share & Ratio\\ 
        \hline \hline
        Computer \& mathematical & 37.2& 3.4&10.9\\  
        Arts, design, sports, entertainment, \& media &10.3&1.4&7.4\\ 
        \textbf{Life, physical, \& social science} & 6.4 &0.9& 7.1\\
        Architecture \& engineering &4.5&1.7&2.6\\
        Educational instruction \& library & 9.3&5.8& 1.6\\ 
        Memo: Other occupations &31.8&87.6&0.4\\ \hline
    \end{tabular}
        \caption*{Note: Percent share of prompts submitted to Claude AI and linked to tasks by Anthropic.  Task weights are apportioned equally to all occupations which include that task in O*NET.\\Source: Anthropic Economic Index.}
\end{table}

\begin{table}
    \centering
    \caption{O*NET Scientific Task Prevalence in Claude AI Prompts}
    \label{table:anthropic2}
    \begin{tabular}{|p{8cm}r|}\hline
        Task & Share (pct.) \\ 
        \hline \hline
        Modelling \& Prediction &  86.5\\\hline
        \footnotesize{conduct logical analyses of business, scientific, engineering, and other technical problems, formulating mathematical \textbf{models} of problems for solution by computers.} & \footnotesize{46.1} \\ \hline
        \footnotesize{design or develop software systems, using scientific analysis and mathematical models to \textbf{predict} and measure outcome and consequences of design.} & \footnotesize{16.9} \\ \hline
        \footnotesize{complete \textbf{models} and simulations, using manual or automated tools, to analyze or \textbf{predict} system performance under different operating conditions.} & \footnotesize{15.7}  \\ \hline
        \footnotesize{develop mathematical or statistical \textbf{models} of phenomena to be used for analysis or for computational simulation.}  & \footnotesize{4.5}  \\  \hline
        \footnotesize{design computer simulations to \textbf{model} physical data so that it can be better understood.} & \footnotesize{2.2}  \\ \hline
        \footnotesize{develop software applications or programming to use for statistical \textbf{modeling} and graphic analysis.} & \footnotesize{1.1}  \\ 
        \hline 
        Other Tasks & 13.5 \\ \hline
        \footnotesize{develop new principles and new relationships between existing mathematical principles to \textbf{advance mathematical science}.}  & \footnotesize{9.0}  \\  \hline
        \footnotesize{\textbf{design research projects} that apply valid scientific techniques and use information obtained from baselines or historical data to structure uncompromised and efficient analyses.} & \footnotesize{4.5}  \\  \hline
    \end{tabular}
    \caption*{Note: Share of all Anthropic prompts accounted for by these tasks is 0.9\%. Task-level prompts labeled by Anthropic.\\Source: Anthropic Economic Index.} 
\end{table}

\Vref{table:anthropic2} shows the prevalence of selected O*NET tasks related to scientific discovery among Claude AI prompts.  These tasks collectively account for only 0.9\% of all prompts, revealing that the share of prompts accounted for by scientists (6.4\%) includes far more than scientific discovery.  These discovery tasks are most commonly related to the creation of mathematical or statistical models of technical phenomena, such as business, scientific, and engineering, either to foster understanding of the phenomena or to predict how modeled systems would perform. Such tasks account for 86.5\% of the scientific tasks Claude AI is asked to help with.  Other tasks include the advancement of mathematical science (9.0\% of scientific prompts) and the design of research projects (4.5\%). 

\begin{figure}
\centering
\caption{GenAI Automation vs. Augmentation in Researcher Roles}\label{figure:anthropic_automation-vs-augmentation}
\includegraphics[width=0.75\textwidth]{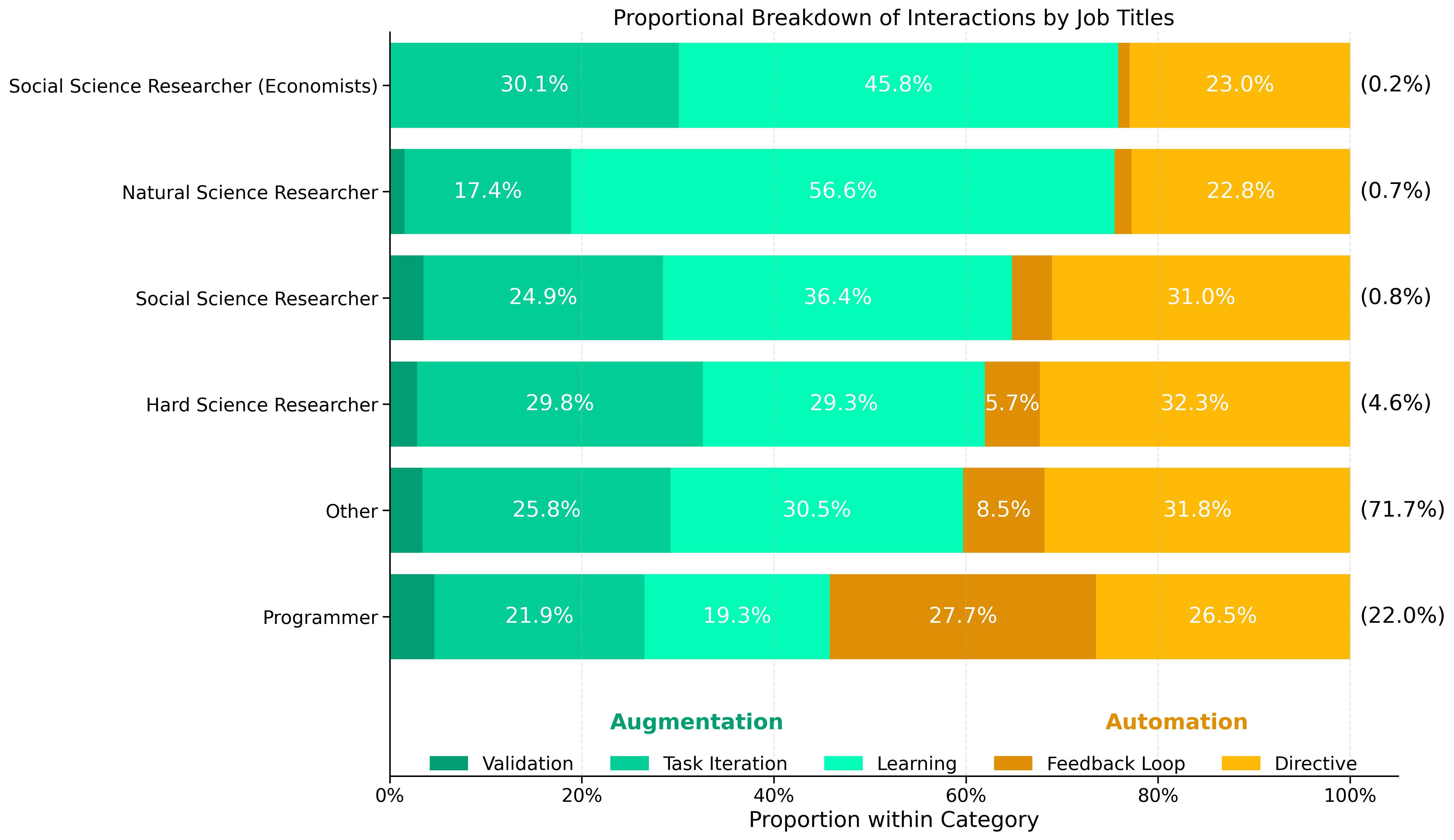}
\begin{minipage}{\linewidth}\setlength{\belowdisplayskip}{0pt} \setlength{\abovedisplayskip}{0pt}
\scriptsize
{\vspace{2em}Note:  Authors' calculations.\\Source: Anthropic Economic Index.}
\end{minipage}
\end{figure}

\Vref{figure:anthropic_automation-vs-augmentation} illustrates significant automation and augmentation of tasks among our groupings of research occupations: programmers exhibit the highest automation rate, with over half of the requests handled by genAI being automation tasks. Social science researchers show slightly lower automation rates, with economists showing over 23\% of their prompts being automation focused. Notably, for hard science researchers (e.g., physicists, biochemists), the share of their genAI use for automation is nearly 15\% higher than their natural science counterparts. This difference likely reflects AI's strength in data-intensive and simulation-based research such as those found in hard sciences like physics and materials science. 

\paragraph{Conference Call Mentions}

\begin{figure}
	\centering
    \caption{Mentions of AI Usage for Research in Conference Calls}\label{fig:calls}
	\includegraphics[width=0.8\linewidth]{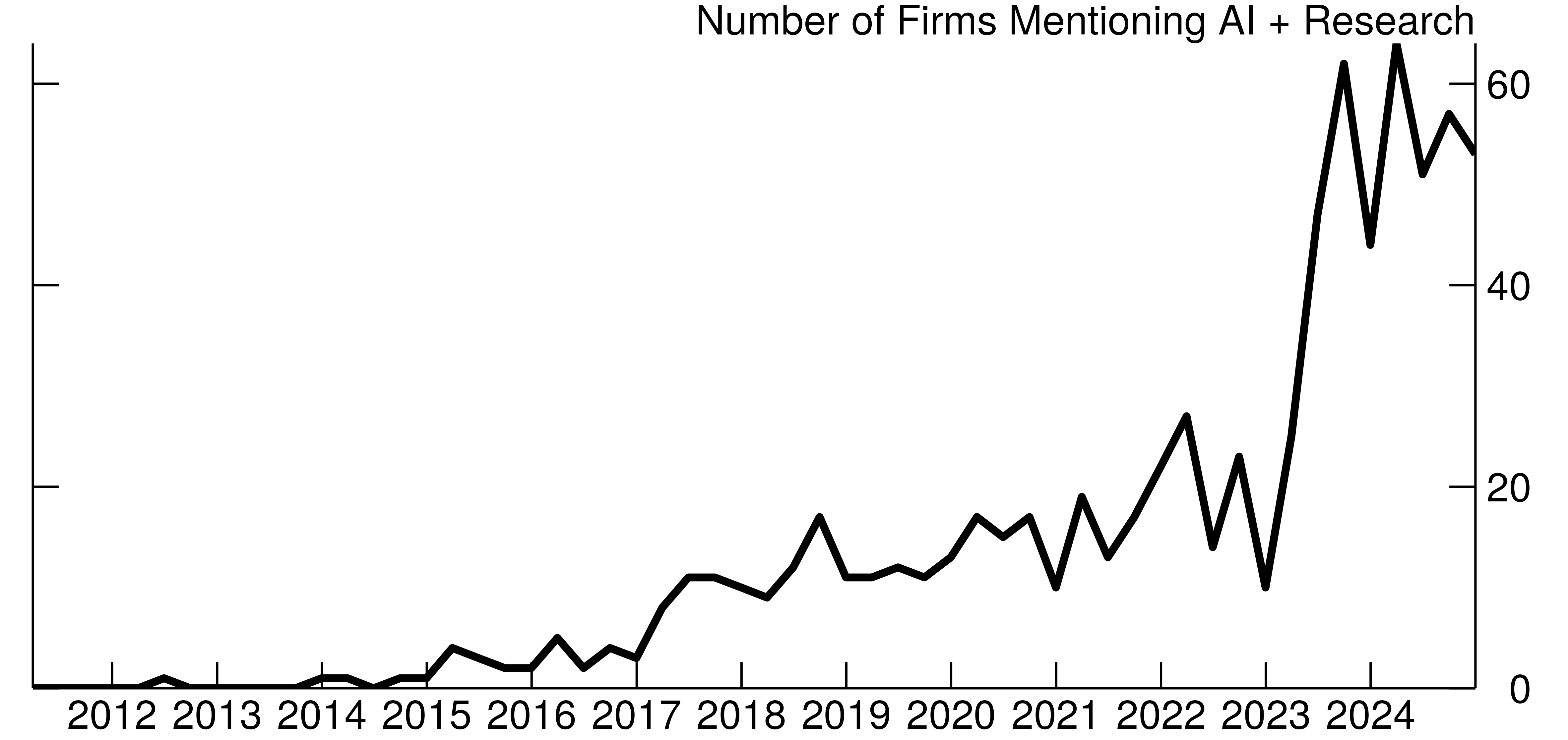}
\begin{minipage}{\linewidth}\setlength{\belowdisplayskip}{0pt} \setlength{\abovedisplayskip}{0pt}
\scriptsize
{\vspace{2em}Note: Authors' estimates.  Data through Q4 2024.\\Source: S\&P Capital database merged with Compustat.}
\end{minipage}
\end{figure}

GenAI's integration into the invention process is also revealed through firm communication. We analyze quarterly earnings calls, which are routine events where firm executives discuss company performance, future projects, and key developments with investors and analysts. \Vref{fig:calls} plots the count of the number of firms referencing AI in the context of research as indicated by the firm mentioning an AI-specific term (``machine learning,'' ``deep learning,'' ``artificial intelligence,'' ``genAI, or ``generative AI'') within a research-related context (within 10 words of ``inventi-'', ``research-'', or ``discover'').  A sudden rise appears in 2023, with approximately 60 public companies per quarter mentioning such usage. This increase in integration of AI with R\&D is illustrative of the role it has begun to play in innovation in a corporate context.  Two examples are provided below.

\begin{samepage}
In 2024, John Wiley \& Sons, a publishing company, announced expansions in compound databases leveraging advanced AI and its capability to accelerate scientific discoveries:\footnote{See John Wiley \& Sons, Inc. FQ1 2025 Earnings Call. \url{https://s27.q4cdn.com/812717746/files/doc_financials/2025/q1/q125-earnings-transcript.pdf}}
\nopagebreak
\vspace{-2mm}
\begin{quote}
Wiley has just released two new database collections using advanced AI techniques to significantly expand the number of compounds available for analysis from food-related compounds to industrial compounds. The end goal here to help scientists reach better conclusions faster.  
\end{quote}
\end{samepage}

\begin{samepage}
Similarly, Cadence Design Systems, an electronic systems design firm highlighted the potential of AI to automate fields such as biology and life sciences.\footnote{See Cadence Deisgn Systems, Inc. FQ3 2023 Earnings Call.}
\nopagebreak
\vspace{-2mm}
\begin{quote}
And then the third phase of AI adoption is AI applied to areas that were not automated in the past, okay? So I think that may take longer, maybe 5 years plus, but that has to be driven to digital biology and life sciences. I mean there's a huge application of AI.
\end{quote} 
\end{samepage}

\subsection{Is GenAI an IMI?}
Our assessment is that there is a strong case that genAI is an IMI.  Indeed, it is a multifaceted IMI, having characteristics of an observational IMI (e.g. image enhancement), an analytical IMI (e.g. sentiment analysis), an organizational IMI (e.g. digital twins), and a communication IMI (e.g. document drafting).  Whether the excitement over genAI research agents will prove to be merited or not, genAI's ability to augment these four dimensions of invention suggests that the idea generation process is becoming more productive.

It also appears that it is taking root within the research community. The signal from the USPTO's AI database is that AI patents surged when the use of genAI became practical.  GenAI prompt analysis from Anthropic points to relatively intense use of AI by scientists as well as in adjacent fields like computing and engineering.  And, corporate earnings calls increasingly mention genAI while discussing their research efforts.

\section{Conclusion}

The release of ChatGPT in late 2022 was a stark inflection point in public interest in genAI and predictions of a first-order impact on productivity in the future soon followed.\footnote{Goldman Sachs analysts forecast that genAI will eventually increase in U.S. labor productivity by 15\%. See Briggs, Joseph, and Sarah Dong. ``Global Economics Comment: AI Productivity and Labor Market Impacts Are Still Small (For Now).'' Goldman Sachs Economic Research, March 14, 2025.}  Yet, as exciting as progress in genAI is from a science and engineering standpoint, its economic effects are highly uncertain.  For firms to justify the reorganization and other complementary capital needed to exploit genAI, the return from the technology, less the total cost of ownership, must be high enough.  Field studies do point to efficiency gains in selected business functions and many firms have experimented with the technology, but only a small share of them attest to material improvements to their bottom lines from the technology thus far.   

To complement the limited empirical evidence, we ask what the characteristics of genAI suggest its future impact on productivity may be.  GenAI has features typical of both a general-purpose technology---headed toward being widely used, stimulating related innovation, and displaying ongoing improvement in (economic) performance---and an invention of a method of invention---raising the efficiency of R\&D through improvements to observation, analysis, communication, or organization.  

Because both GPTs and IMIs promote productivity growth for extended periods, it is reasonable to expect genAI will have a noteworthy impact on productivity. Importantly, genAI's potential for productivity does not depend on the elusive goal of reaching artificial general intelligence (AGI). It can qualify as a GPT and IMI well before the arrival of AGI. The main hurdle is diffusion. Complementary innovations like  interfaces, robotics, and agents, for example, are emerging, and technological progress is ongoing. Yet, outside of the tech sector, firm-level adoption in production processes is still modest. As an IMI, the case is stronger: genAI usage is gaining traction within the scientific community via workflows and patents. Even so, we offer several cautionary observations.  

First, we expect that genAI will boost productivity growth \textit{relative to the counterfactual} economy without it.  If the growth effect of machine learning (and other IT innovations) is waning, the impact of genAI will have to match the impact of machine learning on the likes of Amazon and Facebook for the economy simply to match the recent history of productivity growth.\footnote{\citet{bresnahan:2024:innovation-paths-ai-gpt} notes that the spread of earlier AI technologies to other companies slowed once the digitally native companies had jumped in.}

Second, the GPT effect on productivity is inherently slow as it involves complementary investment. For example, the effect on the productivity \textit{level} of solid-state computing was large, but it played out over decades, damping the effect on productivity \textit{growth}.  The tech boom was a long time coming:  Massive advances in computational technology, including the invention of the solid state transistor and the fundamentals of system design had accumulated by the end of the 1940s and a steady decline in computing costs had begun \citep{nordhaus:2007:two-centuries-productivity-growth-computing}.\footnote{Predictions of an IT-infused future of abundance soon followed, but noteworthy productivity gains only appeared in tandem with time-consuming complementary investment, such as business reorganization. For example, \citet{berkeley:1949:giant-brains-machines-think}, based on observation of the handful of computers in existence, eagerly anticipated automatic address books, libraries, translators, typists, and stenographers, as well as business process optimization, psychological testing and training, weather forecasting, and even weather control.  Others, such as \citet[4]{martin:1960:practical-problems-introducing-computer}, cautioned that productivity gains would be hard-won: ``The data-processing system of an organization is of almost unimaginable complexity.  The introduction of a computer usually involves widespread changes in this complex system.''}  The surge in productivity attributed to information technology arrived some fifty years later.

Third, investment to deploy new technologies is fraught with risk.  If genAI is a widely adopted ``killer app'' that defines a new era of IT, the computing capacity needed to deliver genAI to millions of simultaneous users will be massive.  Anticipation of this outcome helps explain the wave of investment in data centers and electricity generation.  But, building to meet anticipated demand can lead to disastrous consequences, as illustrated by the history of railroad expansion and the associated boom-bust cycles in the \nth{19} Century.\footnote{On the British experience in the 1840s, see \citet{campbell-turner:2012:dispelling-myth-naive-investor-british} and \citet{odlyzko:2012:railway-mania-fraud-disappointed-expectations}.}  For IT systems, the capacity forecasting problem is compounded by the progress of technology, which drives down the hardware investment required to deliver a given level of service.\footnote{This forecasting challenge for fiber optic telecommunications systems, combined with duplicative effort by competing networks, was a major factor behind the economic downturn in 2001 \citep{couper-hejkal-wolman:2003:boom-bust-telecommunications,doms:2004:boom-bust-information-technology-investment}.  On the capital overhang in the telecommunications network in the early 2000s, see \citet[53]{hecht:2016:boom-bubble-bust-fiber-optic}:  “The post-bubble network was vastly overbuilt and underused. In late 2002, consulting firm TeleGeography estimated only 10 percent of the long-haul fibers installed in Europe and North America carried any signals, and that only 10 percent of the wavelengths in those fibers were lit. \ldots Soon, creditors were trying to unload dark-fiber networks for pennies on the dollar.”}  A critical further concern is the availability of electrical power to accommodate the demands of data centers supporting widespread genAI use \citep{pilz-etal:2025:ai-power-requirements-exponential-growth}.  And, of course, R\&D is inherently risky because the returns are erratic:  The chain of choices made between insight from research and greater output per hour is long.

Nevertheless, our forecast for the most likely outcome is for a noteworthy contribution of genAI to the level of labor productivity, though the range of plausible outcomes is wide, with respect to both the magnitude of the total contribution and how that impact is spread over time (hence, the productivity growth rate).

\newpage

\printbibliography

\newpage

\begin{appendices}

\section{Definitions of AI}\label{appendix:definitions-ai}
We illustrate the varied use of the term ``artificial intelligence'' by discussing four influential definitions.  Alan Turing devised a broad, conceptual definition---the ``Turing test''---to determine if a system was indistinguishable from humans in 1950.  The Dartmouth Project, a seminal meeting for the AI field in 1956, provided a more demanding analytical definition with concrete criteria for what AI must do.  In 2020, the  \textit{National Artificial Intelligence Initiative} placed a definition into U.S. law with a rather different set of criteria.  And, since 2020, the U.S. Patent and Trademark Office has used an algorithmic approach: a large language model (LLM) supplemented by reinforcement learning with human feedback (RLHF) that can classify any technology as AI or not based on similarity to descriptions of eight related areas of science and engineering.  Importantly, while the sets of technologies meeting these four definitions have substantial overlap, they are far from identical.  Thus, it is crucial to stipulate the scope of analysis when discussing ``AI'' and its economic effects.\footnote{This set of definitions is far from exhaustive.  See the discussion in \citet{filippucci-etal:2024:impact-artificial-intelligence-productivity-distribution} for a definition of scope for AI usefully grounded in a production function framework as well as references to other definitions.  The OECD, for example, has codified this definition: ``An AI system is a machine-based system that, for explicit or implicit objectives, infers, from the input it receives, how to generate outputs such as predictions, content, recommendations, or decisions that can influence physical or virtual environments. Different AI systems vary in their levels of autonomy and adaptiveness after deployment'' \citep{karine:2024:explanatory-memorandum-updated-oecd-definition}.}

\subsection{Alan Turing}
\citet{turing:1950:computing-machinery-intelligence}, while not offering a definition of artificial intelligence, described a procedure to determine if a machine can imitate human responses well enough that a human interlocutor cannot reliably distinguish between the machine and a human.

This, of course, is the origin of the ``Turing test'' which is often referenced to gauge effective artificial intelligence.  We judge that for most observers, passing the Turing test would be a sufficient condition for a system to be AI, but it isn't a necessary condition in current usage.  After all, few people would be fooled into thinking the machine learning-based recommendation engines used on social media sites are human.  Moreover, mimicking and replacing humans is not the sole goal of the AI field.  Indeed, the adverse social consequences of that approach are a concern of some scholars who recommend a focus on using AI to complement human activity instead.  \citet[14]{brynjolfsson:2022:turing-trap-promise-peril-human} observes:
\begin{quote} We can work on challenges that are easy for machines and hard for humans, rather than hard for machines and easy for humans. The first option offers the opportunity of growing and sharing the economic pie by augmenting the workforce with tools and platforms. The second option risks 
dividing the economic pie among an ever-smaller number of people by creating 
automation that displaces ever-more types of workers.\end{quote}
Another shortcoming of this definition has been raised by philosophers who have disputed the use of the Turing test to assess whether a machine can think. \citet{searle:1980:minds-brains-programs} offers the counterexample (the ``Chinese room argument'') of an individual, ignorant of Chinese, passing the Turing Test by using an instruction manual to connect questions posed in Chinese characters to appropriate responses without understanding their meaning.

\subsection{The Dartmouth Project}
The term ``artificial intelligence'' can be traced to the summer of 1956, when Dartmouth College professor John McCarthy convened the seminal ``Summer Research Project on Artificial Intelligence'' \citep{nilsson:2009:quest-artificial-intelligence}.  The project proposal stated its premise and objectives and contained an implicit definition of AI (in italics) \citep[13]{mccarthy-minsky-rochester-shannon:2006:proposal-dartmouth-summer-research-project}:\footnote{He chose the term ``Artificial Intelligence" to distinguish the field from ``automata theory"---a branch of computer science focused on rule-based mathematical models of computation---and ``cybernetics"---a field focused on control systems, feedback, and communication in machines and living things.}
\begin{quote}
    The study is to proceed on the basis of the conjecture that every aspect of learning or any other feature of intelligence can in principle be so precisely described that a machine can be made to simulate it. An attempt will be made to find how to \textit{make machines use language, form abstractions and concepts, solve kinds of problems now reserved for humans, and improve themselves}.
\end{quote}

The Dartmouth AI definition is far broader than the Turing test.  Systems that pass the Turing test would surely be included in the scope of the Dartmouth definition by virtue of using language, provided a system need not satisfy all the criteria at once.  The project also envisioned systems forming abstractions, anticipating the flexible models found in neural network systems.  That is, the training of these models is agnostic about the analytical structure, rather than calibrating a predetermined functional form.  And, the idea that AI systems will improve themselves hints at the concept of artificial general intelligence.  Last, the ``solve kinds of problems now reserved for humans'' criterion is likely the vernacular definition many casual observers would provide for AI if pressed to do so, though whether present-day observers would reserve the same set of problems for humans as observers in 1956 is unclear.

\subsection{National Artificial Intelligence Initiative}
Naturally, as the topic of AI has become a focus of public policy in recent years, the U.S. legal system has required a definition.  One is provided in the \textit{National Artificial Intelligence Initiative (NAII)} (15 U.S.C. \S\ 9401(3)):
\begin{quote}
The term ``artificial intelligence'' means a machine-based system that can, for a given set of human-defined objectives, make predictions, recommendations or decisions influencing real or virtual environments. Artificial intelligence systems use machine and human-based inputs to
\begin{itemize}
\item[(A)] perceive real and virtual environments;
\item[(B)] abstract such perceptions into models through analysis in an automated manner; and
\item[(C)] use model inference to formulate options for information or action.
\end{itemize}
\end{quote}
Like the Dartmouth definition, this law provides concrete criteria for a system to be AI, but the sets of criteria are rather different.  Unlike the Dartmouth definition, the law requires that systems collect observations from the environment.  Like the Dartmouth definition, though, the NAII definition requires that AI systems form abstractions, and one can loosely compare the ``formulate options for information and action'' criterion to the ``solve problems'' criterion in the Dartmouth definition.  However, there is no mention of language in the government definition, an important part of the Dartmouth and Turing definitions, nor any mention of self improvement.

\subsection{The U.S. Patent and Trademark Office}
The U.S. Patent and Trademark Office (USPTO), identifies patents that ``contain'' AI for the Artificial Intelligence Patent Dataset.  This exercise requires a definition that provides an unequivocal declaration for each patent, unlike the other definitions discussed above.  To that end, the USPTO identified eight ``AI component technologies'' and trained a large language model to identify patents referencing those technologies using an iterative supervised learning process where subject experts identified examples of AI and non-AI for each technology and reviewed the AI algorithm determinations for accuracy \citep[quoted from ][6-8]{giczy-pairolero-toole:2022:identifying-artificial-intelligence-invention-novel}:  

\begin{enumerate}
    \item \textbf{Knowledge processing}: The field of knowledge processing contains methods to represent facts about the world and to derive new facts (or knowledge) from a knowledge base. For example, expert systems generally contain a knowledge base and an inference method to obtain new facts from that knowledge base.
    \item \textbf{Speech}: Speech recognition includes methods to understand a sequence of words given an acoustic signal. For example, the noisy channel model is a statistical approach used to identify the most likely sequence of words given verbal input using Bayes’ rule \citep{russell-norvig:2009:artificial-intelligence-modern-approach}.
    \item \textbf{AI hardware}:  The field of AI hardware includes physical hardware designed to implement artificial intelligence software. For example, Google designed the Tensor Processing Unit (TPU) to run neural network algorithms more efficiently. AI hardware may include logic circuitry, memory, video, processors, and solid-state technologies. It may also include embedded software that implements other AI component technologies, such as machine learning algorithms.
    \item \textbf{Evolutionary computation}:  Evolutionary computation contains a set of computational methods utilizing aspects of nature and, specifically, evolution \citep{russell-norvig:2009:artificial-intelligence-modern-approach}. For example, genetic algorithms include methods for selecting algorithm variants through the selection of optimal random mutations by maximizing fitness.
    \item \textbf{Natural language processing}: Natural language processing contains methods for understanding and using data encoded in human natural language. For example, language models represent probability distributions of language expressions \citep{russell-norvig:2009:artificial-intelligence-modern-approach}.
    \item \textbf{Machine learning}: The field of machine learning contains a broad class of computational learning models. For example, supervised learning classification models are algorithms that learn to classify observations based on pre-labeled training data. Machine learning includes, among other techniques, neural networks, fuzzy logic, adaptive systems, probabilistic networks, regression, and intelligent searching.
    \item \textbf{Computer vision}: The field of computer vision contains methods to extract and understand information from visual input, including images and videos. For example, edge detection identifies the boundaries and borders contained in an image. Additional areas of computer vision include object recognition, manipulation (e.g., transformation, enhancement, or restoration), color processing, and conversion.
    \item \textbf{Planning/control}: The field of planning and control contains methods to identify and execute plans to achieve specified goals. Key aspects of planning include representing actions and states of the world, reasoning about the effects of actions, and efficiently searching over potential plans. Modern control theory includes methods to maximize objectives over time \citep{russell-norvig:2009:artificial-intelligence-modern-approach}. For example, stochastic optimal control considers dynamic optimization in uncertain environments. Additionally, planning and control includes data systems for administration/management (e.g., managing an organization and its employees, including inventory, workflow, forecasting, and time management), adaptive control systems, and models or simulators of systems.
\end{enumerate}

Among the concrete challenges faced by this classification effort are what non-software components of AI systems are included. Tensor Processing Units (TPUs), for example, are chips designed to accelerate AI inference, while graphics processing units (GPUs) were originally designed for graphics rendering but have since been re-engineered to accelerate AI model training.  As challenging as this question is, the inclusion of hardware is essential for a definition that is conceptually stable over time with respect to the function of the technology; the division of tasks between hardware and software varies over time within functionally identical IT systems \citep{hennessy-patterson:2011:computer-architecture-quantitative-approach}.

Further distinction by type of AI would be a welcome refinement.  \citet{cockburn-henderson-stern:2018:impact-artificial-intelligence-innovation-exploratory} found distinguishing among robotic, symbolic, and neural network AI technologies as useful for their work on innovation.  In light of recent developments in the theory and application of AI, identifying patents as related to generative AI would be useful for research as well.

\section{A Short History of AI}\label{appendix:history-ai}

The Dartmouth Summer Research Project, in 1956, is often used as a rough marker of the beginning of the AI field, though the scientists in attendance did not share a theory of what the field entailed \citep{moor:2006:dartmouth-college-artificial-intelligence-conference}.\footnote{For more on the conference, see \citet{nilsson:2009:quest-artificial-intelligence}, \citet{wooldridge:2021:brief-history-artificial-intelligence}, and \citet{olson:2024:supremacy-ai-chatgpt-race-change}.}  And, foundational work took place before the Dartmouth project. For example, \citet{mcculloch:1943:logical-calculus-ideas-imminent-nervous} had studied the use of artificial neurons, \cite{shannon:1948:mathematical-theory-communication} had identified Markov Chains as a potential basis for generating new content, and \citet{turing:1950:computing-machinery-intelligence} had introduced a test for machine intelligence (now known as the Turing test) whereby a human attempts to determine if a hidden interlocutor is a computer or another human.\footnote{Indeed, Andrey Markov identified language as a use for his mathematical structures as early as 1906 \citep{markov:2006:example-statistical-investigation-text-eugene}.}

We sketch subsequent developments in AI theory and application below.  As will be apparent, substantial progress on AI preceded the explosion of attention to AI that followed the introduction of ChatGPT in 2022.

\begin{table}
    \centering
    \caption{Distinguishing Features of AI Models}
    \begin{tabular}{lll}
    \hline
    symbolic & vs. & connectionist    \\
    deterministic & vs. & stochastic    \\
    discriminative & vs. & generative    \\
    \hline
    \end{tabular}
    \label{table:ai-dimensions}
\end{table}

\subsection{Early AI Research\label{section:early-ai-research}}

Following the Dartmouth project, AI research developed models distinguished along several dimensions (\vref{table:ai-dimensions}).  

\begin{itemize}
    \item \textbf{Symbolic AI} encoded a system of explicit rules in computer programs. For example, \citet{newell-shaw-simon:1958:elements-theory-human-problem-solving} designed a model called ``Logic Theorist'' that successfully proved 38 theorems from Whitehead and Russell's \textit{Principia Mathematica} \citep{whitehead-russell:1927:principia-mathematica-vol2}.   \textbf{Connectionist AI}, in contrast, allowed complex rules to emerge organically, sacrificing interpretability for flexibility \citep{james-etal:2017:historical-survey-algorithms-hardware-architectures}. The Perceptron model in \citet{rosenblatt:1958:perceptron-probabilistic-model-information-storage-organization}, foundational for this approach, was a single neuron, used to combine signals from multiple input channels to classify images.  Later models, such as MADALINE (Multiple Adaptive Linear Neuron), combined multiple layers of neurons together.  These networks combine input signals into an output signal with the weight given to each input signal evolving during the training process. 
    \item Though early models were typically \textbf{deterministic AI} systems, where a given input would consistently produce the same output, \textbf{stochastic AI} models emerged as well where the path taken was not predetermined.  For example, SNARC (Stochastic Neural Analogue Reinforcement Calculator),  simulated a rat navigating a maze by random experimentation with different paths \citep{minsky:1952:neural-analogue-calculator-probability-model}.
    \item Most early efforts focused on classification---\textbf{discriminative AI}---such as using the Perceptron to label pictures.  But, the ambition to create a \textbf{generative AI} system that would respond to questions with an appropriate free-form text response was already present in this era.  In 1966, the ELIZA chatbot provided a rudimentary simulation of a conversation with a psychotherapist. Unlike present-day AI chatbots, ELIZA used a deterministic, symbolic logic approach, relying on pattern matching and word substitution \citep{weizenbaum:1966:eliza-computer-program-study-natural}.\footnote{Strictly speaking, some AI models, such as the ``expert systems'' described below, are neither generative or discriminative, so our classification scheme is not exhaustive.}
    \end{itemize}

In addition to these theoretical characteristics, applied AI systems are distinguished by the type of training used in their development.  Some use \textbf{reinforcement learning}, interacting with the environment to refine the model.  Others use \textbf{predictive learning}, where the system is trained in advance of use.  Predictive learning primarily took the form of \textbf{supervised learning} in this period, such as when the Perceptron was trained with labeled pictures.  However, \citet{selfridge:1958:pandemonium-paradigm-learning} was a major advance in algorithms for pattern recognition, which was foundational for \textbf{unsupervised learning}, where the system develops classification categories without guidance.

Early efforts to apply theoretical models were limited by advances in computing hardware. Greater AI system capability typically requires a larger model, where size is measured in the number of parameters (\vref{figure:number-parameters-training-dataset-size}). Early models, like Theseus---a robotic maze-solving mouse---and the Perceptron---the rudimentary neuron mentioned above---had tens or hundreds of parameters. Recent models, like DALL-E, Llama, and GPT-3 have hundreds of billions of parameters.  Moreover, more complicated models typically require larger training datasets (\vref{figure:number-parameters-training-dataset-size}).  Computational requirements for model training and application rise with the size of the model and the size of the training dataset. 

\subsection{Emergence of Practical AI}

Early practical applications of AI were found in solving classification problems in high-volume communication systems. \citet{lecun:1989:backpropagation-applied-handwritten-zip-code} developed the LeNet model adopted by the U.S. Postal Service to read hand-written ZIP codes. The post office was soon reading entire handwritten addresses using AI as well \citep{srihari:1997:integration-hand-written-address-interpretation}.  Another early practical application of AI was the identification of spam email \citep{sahami:1998:bayesian-approach-filtering-junk-mail}.  

Notwithstanding these advances in the 1980s and 1990s, interest in the connectionist approach, and neural networks in particular, had fallen off with the downbeat assessment of \citet{minsky-papert:1969:perceptrons}.  Interest returned with the insights provided by \citet{hinton-salakhutdinov:2006:reducing-dimensionality-data-neural-networks}, who introduced advances in training methods (greedy layer-wise pre-training) and efficient use of large datasets (dimensionality reduction).   A key innovation that followed soon after was the \textbf{convolutional neural network} (CNN), an advance in connectionist AI that focused on rapid development of a coarse representation of the image which revealed some features, like edges, but not others.  \citet{krizhevsky-sutskever-hinton:2012:imagenet-classification-deep-convolutional-neural} demonstrated the power of CNN by leaping ahead of other competitors in the \textit{ImageNet Large Scale Visual Recognition Challenge}, a benchmark for computer vision, with their \textit{AlexNet} system.  Referring to the characteristics described above, these early image systems were connectionist, deterministic, discriminative, and trained by predictive learning.  \textit{AlexNet} also demonstrated the value of data augmentation by adding mirror images of training pictures to the training set. 

In the news industry in this period, symbolic models such as \textit{Cyborg} at Bloomberg and \textit{Heliograf} at the Washington Post were used to write articles. Unlike present-day generative models, these systems relied on structured data---tagged as sports scores or stock prices, for example---not on models of the language as a whole, making them a kind of proto-generative AI.   In 2013, articles on major company financial announcements and sports events were generated with these systems and by 2017, these models were used for expanded coverage of sparsely populated news markets and small companies, and were even used to generate rudimentary news videos \citep{keohane:2017:news-writing-bots-future-journalism}. 

Symbolic AI was put to practical use in this period as well.  \textbf{Expert systems} leveraged a large trove of domain-specific information and a set of rules (encoded in an ``inference engine'') provided by specialists to provide guidance, such as medical diagnoses \citep{buchanan:1988:fundamentals-expert-systems}.  Examples include MYCIN, used to diagnose infectious diseases, and IBM's Watson, deployed in medical and other applications \citep{shortliffe:1977:mycin-knowledge-based-computer-program, ferrucci:2012:introduction-this-watson}.  Expert systems fell out of favor over time due to their cost of development, limited reliability, and narrow field of application \citep{gill:1995:early-expert-systems-where-now}.

Most importantly, as emphasized by \citet[32]{agrawal-gans-golfarb:2019:artificial-intelligence-ambiguous-labor-market}, AI practitioners soon realized that discriminative AI could be recast as ``prediction in the statistical sense of using existing data to fill in missing information'' and these models were soon used in a diverse set of prediction problems.  (Indeed, these systems are often now referred to as ``predictive AI.'') Amazon, for example, first used AI to forecast demand for its products in 2009, then continuously updated its forecasting approach to adopt emerging AI techniques (\cref{table:amazon-forecasts}).  \textbf{Machine learning}, another term for this approach to AI, is credited by Amazon and other major IT companies with increasing profitability during this period \citep{bresnahan:2019:artificial-intelligence-aggregate-growth-prospects}.

\begin{table}
    \centering
    \caption{History of Amazon Demand Forecasting Techniques}
    \label{table:amazon-forecasts}
    \begin{tabular}{|c|l|}
    \hline
         Year& Forecasting Technique \\ \hline \hline
         2007 & Time-Series Models\\
         2009 & Random Forest \\
         2011 & Seasonality Models \\
        2013 & Sparse Quantile Random Forest \\
         2015 & Feed-Forward Networks \\
         2017 & Multi-Horizon Quantile Recurrent Forecaster\\
         2020 & MQ Transformer\\ \hline \hline
             \multicolumn{2}{|l|}{Source: Reproduced from \citet{hardesty:2019:history-amazon-recommendation-algorithm}}\\ \hline
    \end{tabular}
\end{table}

\subsection{Generative AI \label{section:generative-ai}}

Public interest in AI surged in late 2022 with the appearance of ChatGPT, a user interface for a viable genAI system---one that can respond to natural language prompts with human-like (coherent, nuanced, context-specific) responses in the form of text, images, videos, and sounds.  The event was the culmination of a roughly 10-year period of advances in \textbf{large language models} (LLMs).   

LLMs are a mathematical representation of the linguistic relationships among the ``tokens'' (words, groups of words, and portions of words) found in a ``corpus'' (set of texts or other media).  A key advantage of LLMs is their ability to reduce unstructured text to flexible structural representations that do not rely on a small set of variables specified in advance.  Rudimentary early attempts at language models encoded words as long vectors of zeros with a single element---the index assigned to the specific word---marked with a ``1'' \citep{hancock-khoshgoftaar:2020:survey-categorical-data-neural-networks}. In 2013, Google introduced a richer method known as the Word2Vec model \citep{mikolov-sutskever-le:2013:learning-meaning-behind-words}. Words are represented with dense vectors known as \textbf{embeddings} such that the distance between two embeddings reflects the semantic similarity between the represented words.  Word2Vec revolutionized many \textbf{natural language processing} (NLP) tasks, such as classification and translation.  

A limitation of Word2Vec encodings is that a word's representation is the same regardless of the context. For example, the model will represent the word ``bank" with the same embedding whether the input text is ``I withdrew money from my bank account" or ``I went fishing down at the river's bank." Since Word2Vec assigns fixed embeddings, it cannot distinguish whether ``bank" refers to a financial institution or a riverbank, impeding its understanding of the text.\footnote{Computer scientists have wrestled with this word sense disambiguation problem since the 1950s. \citet{bar-hillel:1960:demonstration-nonfeasibility-fully-automatic-high} in discussing the prospects for fully automatic high-quality translation, offered this assessment: ``What such a suggestion amounts to, if taken seriously, is the requirement that a translation machine should not only be supplied with a dictionary but also with a universal encyclopedia. This is surely utterly chimerical
and hardly deserves any further discussion.''  It appears we now have such a universal encyclopedia in hand.} 

A major breakthrough in addressing this shortcoming came with the introduction of the Transformer model, an architecture that creates context-aware word representations by efficiently processing the entire input text at once \citep{vaswani-shazeer-parmar:2017:attention-all-you-need}. This architecture is defined by two principal characteristics: the \textbf{attention mechanism} and \textbf{positional encodings}. The attention mechanism enables the model to assign, for each token, varying degrees of relevance to different parts of the input, allowing it to understand context in longer sequences of text. Positional encodings ensure that word order is meaningfully integrated into the model's processing of inputs. (See the box, ``Landmark AI Models: The Transformer,'' for more detail.) This advancement changed how machines encoded text, shifting the focus within NLP toward a deeper understanding of language.\footnote{Particularly important was the introduction of the BERT model the following year \citep{devlin:2018:bert-pre-training-deep-bidirectional}. The final `T' in BERT stands for `Transformer' (bidirectional encoder representations from transformers)} 

With the advent of the Transformer, genAI flourished. Most notably, text generation improved at a blistering pace in this period, though NVIDIA's generative adversarial network (GAN) (2018) and DALL-E (2021) were striking advances in image generation; audio generation was dominated by WaveNet (2016), a different architecture, for a time, but eventually the Transformer approach was used for speech generation as well. Especially prominent were a series of models produced by OpenAI: GPT-2 \citep{radford-etal:2019:language-models-unsupervised-multitask-learners}, GPT-3 \citep{brown-etal:2020:language-models-few-shot-learners}, ChatGPT \citep{openai:2022:introducing-chatgpt}, GPT-4 \citep{openai:2023:gpt4-technical-report}, and others.  Successive models were increasingly human-like in their knowledge and creativity, eventually passing Turing tests due to their coherent and contextually relevant output.  

\end{appendices}
\end{document}